\documentclass[a4paper,twocolumn,unpublished]{quantumarticle}
\pdfoutput=1

\usepackage[utf8]{inputenc}
\usepackage[english]{babel}
\usepackage[T1]{fontenc}
\usepackage{booktabs}
\usepackage{graphicx}
\usepackage{latexsym}
\usepackage{amssymb}
\usepackage{amsmath}
\usepackage{amsthm}
\usepackage{mathdots}
\usepackage{amsfonts}
\usepackage{bm}
\usepackage{multirow}
\usepackage{color}
\usepackage{comment}
\usepackage{dsfont} % provides \mathds as replacement for \mathbbm
\usepackage{marvosym}
\usepackage{wasysym}
\usepackage{enumerate}  
\usepackage{hyperref}
\usepackage{url}
\usepackage{caption}
\usepackage[table]{xcolor}

\newcommand{\lrar}{\leftrightarrow}
\newcommand{\Z}{\mathbb{Z}}

\newcommand{\prob}{\mathrm{prob}}

\definecolor{cinnabar}{rgb}{0.89, 0.26, 0.2}

\newcounter{para}

\usepackage{array, makecell}

\begin{document}
\title{Rigorous estimation of error thresholds of transversal Clifford logical circuits}
\author{Yichen Xu}
\affiliation{Department of Physics, Cornell University, Ithaca, NY, USA}
\email{yx639@cornell.edu}
\author{Yiqing Zhou}
\affiliation{Department of Physics, Cornell University, Ithaca, NY, USA}

\author{James P. Sethna}
\affiliation{Department of Physics, Cornell University, Ithaca, NY, USA}

\author{Eun-Ah Kim}
\affiliation{Department of Physics, Cornell University, Ithaca, NY, USA}

\maketitle

\begin{abstract}
The threshold theorem promises a path to fault-tolerant quantum computation, provided the physical error rate is below a critical threshold. While transversal gates efficiently implement logical operations, they propagate errors and can lower this threshold relative to a static quantum memory. 
%Existing threshold estimates for transversal circuits are empirical and restricted to specific, sub-optimal decoders.
In this work, we generalize the statistical-mechanical (stat-mech) mapping from quantum memories to logical circuits with transversal gates, thereby enabling rigorous, decoder-independent thresholds for fault-tolerant logical computation. We first demonstrate the framework for two toric code blocks undergoing a transversal CNOT (tCNOT) gate, quantifying two independent error-spreading mechanisms. For persistent bit-flip errors with perfect syndromes, the stat-mech model is a 2D random Ashkin-Teller model. Monte Carlo simulation and finite-size scaling show that the tCNOT reduces the optimal bit-flip threshold to $p=0.080$, a $26\%$ decrease from the toric code memory threshold $p=0.109$. With syndrome errors included, the circuit maps to a 3D random 4-body Ising model with a plane defect, yielding a conservative estimate $p\geq 0.028$, a modest $15\%$ reduction from the memory threshold $p=0.033$. Beyond the tCNOT gate, we derive stat-mech models for all transversal Clifford gates of the toric code, including the fold-transversal Hadamard and $S$ gates, and generalize the framework to arbitrary CSS codes, proving that each transversal gate modifies the stat-mech model only locally in time.
By reducing threshold analysis of fault-tolerant logical circuits to the study of classical spin models with local defects, our framework provides a systematic, decoder-independent benchmark for near-term fault-tolerant architectures.
\end{abstract}

\section{Introduction}

The threshold theorem~\cite{aharonov1997Proc.Twenty-NinthAnnu.ACMSymp.TheoryComput.-STOC97,kitaev1997Russ.Math.Surv.,knill1998Proc.R.Soc.Lond.A,aliferis2005} is a cornerstone of quantum error correction (QEC), promising a path toward fault-tolerant quantum computation (FTQC)~\cite{shor1995Phys.Rev.Aa,preskill1997,gottesman1997b}. The theorem states that for a quantum device with a physical error rate below a non-zero threshold, a QEC scheme can suppress the logical error rate to an arbitrarily low level. For instance, the toric code used as a quantum memory has a well-established bit-flip error threshold of approximately 10.9\% for perfect syndrome measurements, or 3.3\% when including syndrome errors~\cite{dennis2002J.Math.Phys.a,wang2003AnnalsofPhysics,ohno2004NuclearPhysicsBb}. Encouragingly, recent hardware breakthroughs have pushed various quantum platforms into this sub-threshold regime, enabling demonstrations where error correction improves the fidelity of logical memory~\cite{acharya2023Natureb,acharya2025Nature}. These advances set the stage for the implementation of fault-tolerant logical computations~\cite{bluvstein2024Naturec,reichardt2025,lacroix2025Nature,paetznick2024,ryan-anderson2024science}. Two dominant schemes for this task are transversal gates~\cite{gottesman1998Phys.Rev.A,knill2005Nature,bombin2006Phys.Rev.Lett.a} and lattice surgery~\cite{horsman2012NewJ.Phys.b}, each having its own pros and cons. Specifically, compared with lattice surgery, transversal gates require fewer physical qubits and a shallower circuit. However, physical errors propagate across code blocks via transversal gate operations, thereby inevitably lowering the error threshold. Rigorously establishing how transversal gates affect the error threshold has thus become a pressing question for the pursuit of fault-tolerant computation.

Although Ref.~\cite{zhou2025Naturea} proved the existence of a non-zero error threshold for transversal logical circuits, the actual threshold values have only been observed empirically for specific, practical decoders~\cite{Beverland2021cost,cain2024Phys.Rev.Lett.a,wan2025,sahay2025PRXQuantum,cain2025a,serra-peralta2025,turner2025,zhou2025}. Decoder-specific thresholds underestimate the fundamental limit: different decoders yield different values, and none provides the decoder-agnostic \textit{optimal} threshold that bounds the performance of any decoding strategy. For quantum memories, the optimal threshold is established through the statistical mechanical (stat-mech) mapping~\cite{dennis2002J.Math.Phys.a,wang2003AnnalsofPhysics}, a powerful analytical framework that maps the physical error rate to temperature and disorder strength of a classical spin model, and identifies the threshold with the critical point of an order-disorder phase transition. This mapping leverages the correspondence between the large-code-distance limit and the thermodynamic limit of the classical model, enabling well-controlled finite-size scaling (FSS) analysis to pinpoint thresholds with high precision. The stat-mech mapping has been successfully applied to a variety of topological codes and error models (see Table~\ref{tab:survey}), in every case yielding insights inaccessible to decoder-based empirical studies.

Despite this considerable body of work on quantum memories, the stat-mech mapping has not yet been extended to logical circuits involving transversal gates. All existing stat-mech mappings listed in Table~\ref{tab:survey} are restricted to static quantum memories without logical operations. The fundamental obstacle is that transversal gates dynamically propagate errors between code blocks, modifying the structure of undetectable error cycles in a way that has no analog in the memory setting.
In principle, the optimal threshold of a transversal logical circuit can be crudely estimated by circuit simulation with a most likely error (MLE) decoder~\cite{cain2024Phys.Rev.Lett.a,zhou2025Naturea}, due to its near-optimal performance for the toric code. However, since finding the MLE is in general NP-hard for logical circuits, such an empirical approach is limited to small code distances, precluding controlled extrapolation to the thermodynamic limit.

In this work, we overcome this obstacle and generalize the stat-mech mapping to logical circuits with transversal Clifford gates. We show that the transversal gate modifies the classical spin model only locally in time, introducing a permutation defect of the Ising spins at the time slice where the gate is applied. This structural insight enables, for the first time in the context of logical circuits, rigorous, decoder-independent threshold analysis via FSS. In the main text, we demonstrate this framework for two toric code blocks undergoing a transversal CNOT (tCNOT) gate, which serves as a representative example for entangling transversal Clifford gates. We consider two complementary error models. First, for persistent bit-flip errors with perfect syndromes, we map the decoding problem to a 2D random Ashkin-Teller (AT) model, and use Monte Carlo simulations and FSS to determine that the tCNOT gate reduces the optimal bit-flip error threshold of the target code block to $p=0.080$, a $26\%$ decrease from the memory threshold. Second, for bit-flip errors coexisting with syndrome errors, the logical circuit maps to a 3D random 4-body Ising model with a plane defect; leveraging previously reported numerical results~\cite{ohno2004NuclearPhysicsBb}, we conservatively estimate a threshold of $p=0.028$, implying a more modest $15\%$ reduction from the memory threshold.

This paper is organized as follows. In Sec.~\ref{sec:prelim}, we review the stat-mech mapping for toric code quantum memory. In Sec.~\ref{sec:statmech}, we introduce the tCNOT gate and derive the stat-mech model for persistent bit-flip errors with perfect syndromes, followed by Monte Carlo simulation to determine the corresponding error thresholds. In Sec.~\ref{sec:noisysyn}, we derive the stat-mech model with syndrome errors and estimate the threshold reduction. Finally, in Sec.~\ref{sec:conclusion}, we discuss the implications of our stat-mech framework for fault-tolerant quantum computation.

Going beyond the tCNOT gate, in the Appendices we complete the stat-mech mapping for all transversal Clifford gates of the toric code, including the fold-transversal Hadamard and $S$ gates (Appendix~\ref{app:foldtransversal}), and further generalize the framework to arbitrary Calderbank-Steane-Shor (CSS) codes with transversal Clifford gates using a binary matrix formalism (Appendix~\ref{app:general}). In the most general setting, we prove that each transversal gate locally modifies the stat-mech model at the time it is implemented, thereby establishing the scalability of our approach to entire logical circuits.

\begin{table*}
    \centering
    \small
    \newcolumntype{C}[1]{>{\centering\arraybackslash}p{#1}}
    \begin{tabular}{C{1.1cm}C{2.8cm}C{2.6cm}C{1.2cm}C{5.5cm}}\toprule
       \makecell{Logical \\ ops?} & Stabilizer code  & Error model & \makecell{Synd. \\ error?} & Stat-mech model \\ \midrule \multirow{11}{*}{\makecell{No}} & \multirow{6}{*}{\makecell{2D toric \\ code}} & \multirow{2}{*}{\makecell{Bit/phase \\ flip}}  & No & 2D RBIM ($p_\text{th}\!\approx\! 0.109$) \cite{dennis2002J.Math.Phys.a} \\\cmidrule(lr){4-5}
         & & & Yes & 3D R4bIM ($p^*\!\approx\! 0.033$) \cite{wang2003AnnalsofPhysics,ohno2004NuclearPhysicsBb}\\ \cmidrule(lr){3-5}
         & & Depolarizing & No & 2D Random AT model \cite{bombin2012Phys.Rev.X}\\\cmidrule(lr){3-5}
         & & Correlated & No & RBIM + further neighbor \cite{chubb2021Ann.Inst.HenriPoincareComb.Phys.Interact.}\\\cmidrule(lr){3-5} 
         & & \makecell{Single-qubit \\ coherent} & No &  \multirow{2}{*}{\makecell{Random complex \\ coupling AT \cite{behrends2024,behrends2025}}}\\ \cmidrule(lr){3-4}
         & & \makecell{Coherent + \\ incoherent} & No &  \\ \cmidrule(lr){2-5}
         & \multirow{2}{*}{\makecell{2D color \\ code}} & \multirow{2}{*}{\makecell{Bit/phase \\ flip}} & No & 2D R3bIM \cite{bombin2008Phys.Rev.A,katzgraber2009Phys.Rev.Lett.}\\ \cmidrule(lr){4-5}
         & & & Yes & 3D Random Ising gauge theory \cite{andrist2011NewJ.Phys.}\\ \cmidrule(lr){2-5}
         & 3D toric code & Bit/phase flip & No & 3D RBIM/R4bIM \cite{ohno2004NuclearPhysicsBb,hasenbusch2007Phys.Rev.B}\footnote{The 3D toric code and color code are not self-dual, hence the stat-mech models of bit and phase flip errors are different.}\\\cmidrule(lr){2-5}
         & 3D color code & Bit/phase flip & No & 3D R4bIM/R6bIM \cite{kubica2018Phys.Rev.Lett.}\\\hline
        \multirow{2}{*}{\makecell{Yes}} & \multirow{2}{*}{\makecell{2D toric code \\ + tCNOT \\ \textbf{(this work)}} }  & \makecell{Persistent \\ bit/phase flip} & No & 2D Random AT model\\ \cmidrule(lr){3-5}
         & & Bit/phase flip & Yes & 3D R4bIM + plane defect\\ \cmidrule(lr){2-5} 
         & \makecell{CSS code + \\ transv.\ Clifford} & Generic Pauli & Yes & See Appendix~\ref{app:general} \\
        \bottomrule
    \end{tabular}
    \caption{Stat-mech mappings for QEC codes. All prior works (top section) apply exclusively to static quantum memories without logical operations. This work (bottom section) provides the first stat-mech mappings for logical circuits with transversal gates. RBIM: random bond Ising model. AT: Ashkin-Teller. R$n$bIM ($n=3,4,6$): random $n$-body Ising model. 3D R4bIM is also known in the literature as the random plaquette Ising gauge theory. }
    \label{tab:survey}
\end{table*}

\section{Stat-mech mapping for error threshold of quantum memories}
\label{sec:prelim}
   Determining the error threshold via a stat-mech mapping comprises two steps. The first step is to map the probability distribution of syndromes to the partition function of a stat-mech model. The second step is to identify the thermodynamic quantity in the stat-mech model that corresponds to the logical error rate. Since our derivation closely follows the derivations of quantum memories, we first review the stat-mech mapping for quantum memories for completeness, following Refs.~\cite{dennis2002J.Math.Phys.a,wang2003AnnalsofPhysics,chubb2021Ann.Inst.HenriPoincareComb.Phys.Interact.}.

\begin{figure}
    \centering
    \includegraphics[width=0.65\linewidth, page=2]{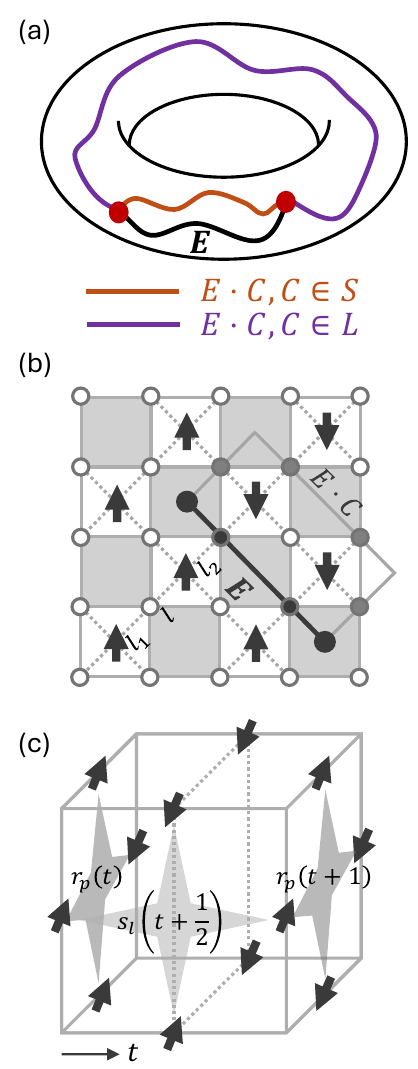}
    \caption{Stat-mech mapping of toric code under errors. (a) Sketch of a trivial $C\in S$ and a non-trivial cycle $C\in L$ on top of an error $E$ in one toric code block, which triggers a pair of syndromes denoted by the two red dots. (b) Stat-mech model of toric code under bit-flip errors. Here, the physical qubits of the toric code are marked by circles, and the weight-4 Z stabilizers are defined on the shaded plaquettes. A trivial cycle $C\in S$ can be parameterized by a domain wall of Ising spins $\{\sigma\}$ (marked by the arrows) on a 2D square lattice that is rotated by $45^\circ$, marked by dashed lines. Two ends of a link $l$ on this lattice are denoted by $l_{1,2}$.  (c) The 3D stat-mech model for syndrome errors in a single code block. The Ising spins reside on the links of the 3D cubic lattice, as indicated by the arrows. Some of the 4-body interaction terms in the Hamiltonian in Eq.~\eqref{eq:R4bIM} are marked by grey stars. The dashed lines at $t\in\mathbb{Z}+\frac{1}{2}$ form a 2D square lattice where Ising spins live on the sites, which is the same 2D square lattice in (b). }
    \label{fig:statmech}
\end{figure}

\subsection{Stat-mech mapping with perfect syndromes}
   We first consider the case of perfect syndromes, in which every stabilizer measurement returns the actual physical value of the stabilizer. In this case, the error in the code state consists solely of physical Pauli errors, which we denote by the Pauli operator $E$. The error model is defined as the probability distribution $\prob(E)$. Physically, the Pauli error is a quantum channel that maps the density matrix of the system $\rho$ to $\sum_E \prob(E) E\rho E^\dagger$. In this case, the logical error rate, $P_\text{logical}$, can be computed as follows~\cite{dennis2002J.Math.Phys.a}:
\begin{align}\label{eq:logicalprob}
    P_\text{logical}&=\sum_E \prob(E) \left[\sum_{C\in L} \prob(E \cdot C|E)\right]\nonumber\\
    &=\sum_E \prob(E)\frac{\sum_{C\in L} \prob(E \cdot C)}{\sum_{C\in S\cup L} \prob(E \cdot C)},
\end{align}
where $C$ denotes a ``cycle" of Pauli operators that do not trigger any syndromes when the stabilizers are measured\footnote{We call $C$ a cycle since in 2D topological codes such an operator is usually supported on closed cycles.}. That is, it either belongs to the stabilizer group of the code state, denoted by $S$, or the set of non-trivial logical operators, denoted by $L$. Thus, Pauli errors $E$ and $E\cdot C$ will trigger the same set of syndromes.
In the first equality, the expression in square brackets is the probability of a logical error given that a physical error $E$ occurs, which is computed via the second equality. 
Without the knowledge of the ground truth about $E$,  the optimal decoder guesses an error $E\cdot C$ according to the conditional probability $\prob(E\cdot C|E)$.
A logical error will occur whenever our guess differs from the actual physical error by any logical operator, see Fig.~\ref{fig:statmech}(a).
Averaging over all possible errors yields the logical error rate in Eq.~\eqref{eq:logicalprob}.

   The central step of stat-mech mapping is to rewrite the summation of the cycles $C$ in Eq.~\eqref{eq:logicalprob} into a partition function of a stat-mech model with quenched disorder. 
For the toric code under a single qubit bit flip error, the cycle $C$ can be parameterized using a set of Ising spins $\{\sigma\}$ living on the lattice sites of a 2D square lattice, see Fig. \ref{fig:statmech}(b). A topologically trivial cycle $C\in S$ is mapped to a topologically trivial domain wall of Ising spins.  The summation of $C\in S$ now becomes the following partition function:
\begin{align}\label{eq:sumc}
    &\sum_{C \in S} \prob(E\cdot C)\propto Z(E)=\sum_{\{\sigma\}}e^{-H(\{\sigma\}|E)},\nonumber\\
    &H(\{\sigma\}|E)=-J\sum_ls_l\sigma_{l_1}\sigma_{l_2},
\end{align}
where $H(\{\sigma\}|E)$ is the Hamiltonian of the 2D random bond Ising model (RBIM). Here $J=\frac{1}{2}\ln\frac{1-p}{p}$ is related to the bit-flip error rate $p$, $l_{1,2}$ are two lattice sites at the ends of the link $l$, and the sign of the random bond interaction, $s_l=1/-1$, corresponds to the absence/existence of physical errors of the qubit that is located at the link $l$. 

   Meanwhile, the sum over the non-trivial cycles $C\in L$ becomes the sum of partition functions with all possible non-trivial global domain walls (DW) in the Hamiltonian, i.e.
\begin{equation}
    \sum_{C\in L_a}\prob(E\cdot C)\propto Z(E)\equiv \sum _{\{\sigma\}}e^{-H_{DW_a}(\{\sigma\}|E)},
\end{equation}
where $a=1,2,3$ denotes the three topologically non-trivial global domain wall configurations around the 2D torus, each giving rise to a different logical error. Therefore, the logical error rate in terms of the stat-mech model is
\begin{equation}\label{eq:logicalerror}
    P_\text{logical}=\sum_E \prob(E)\sum_a e^{-F_{DW_a}(E)} =\sum_a \left\langle{e^{-F_{DW_a}}}\right\rangle,
\end{equation}
where $F_{DW_a}(E)\equiv -\ln\frac{\sum _{\{\sigma\}}e^{-H_{DW_a}(\{\sigma\}|E)}}{\sum _{\{\sigma\}}e^{-H(\{\sigma\}|E)}}$ is the domain wall free energy of a global domain wall in the topological class $a$, and $\langle\cdots\rangle$ denotes the disorder average over the error configuration $E$. 

   The error threshold now has a clear physical picture: the critical point of the order-disorder phase transition of the stat-mech model. For low error rate, the stat-mech model has a low effective temperature (i.e. the couplings in $H(\{\sigma\}|E)$ are strong), and the stat-mech model is in the ordered phase. Consequently, the domain wall free energy $F_{DW}$ scales with the system size. Therefore, the logical error rate is exponentially suppressed as the code distance increases. In contrast, the high error rate is mapped to the disordered phase of the stat-mech model at high temperature, where the domain wall free energy $F_{DW}$ is finite. Therefore, $P_\text{logical}$ remains finite even in the large code distance limit. Based on numerical simulations in Ref.~\cite{honecker2001Phys.Rev.Lett.a}, the error threshold is $p_\text{th}=0.109$.

\subsection{Stat-mech mapping with syndrome errors}
\label{sec:statsynderror}
   We then review the stat-mech mapping for syndrome errors. 
A common way to model this is to assume that every stabilizer measurement has the probability $q$ of being the opposite of its true physical value. In this case, one generally needs multiple rounds of syndrome extractions to suppress logical errors. Therefore, the corresponding stat-mech model becomes three-dimensional~\cite {wang2003AnnalsofPhysics}. Again, the core in the stat-mech mapping procedure is to represent the summations over $C$ in Eq.~\eqref{eq:logicalprob} in partition functions.
In the case of toric code memory under bit-flip noise with syndrome errors,  every cycle $C$ can be parameterized by a set of Ising spins $\{\sigma\}$, which lives on the links of the 3D cubic lattice, see Fig.~\ref{fig:statmech}(c). 
The corresponding stat-mech model is the 3D random 4-body Ising model(R4bIM), also known as the random plaquette Ising gauge theory~\cite{dennis2002J.Math.Phys.a,wang2003AnnalsofPhysics,ohno2004NuclearPhysicsBb}:
\begin{align}\label{eq:R4bIM}
    &\sum_C \prob(E\cdot C)\propto Z_{3D}(E)=\sum_{\{\sigma\}}e^{-H_{3D}(\{\sigma\}|E)},\nonumber \\
    &H_{3D}(\{\sigma \}|E)=-K\sum_{p,t}r_{p}(t)\prod_{l\in p}\sigma_l(t)\nonumber\\
    &-J\sum_{l,t}s_l(t+\frac{1}{2})\sigma_l(t)\sigma_l(t+1)\sigma_{l_1}(t+\frac{1}{2})\sigma_{l_2}(t+\frac{1}{2}).
\end{align}
Here, the coupling constants are related to the error rates via $J=\frac{1}{2}\ln\frac{1-p}{p}$ and $K=\frac{1}{2}\ln\frac{1-q}{q}$. The error $E$ is parameterized by two sets of binary numbers: $s_{l}(t+\frac{1}{2})=1/-1$ for the absence/occurrence of a physical error at the data qubit on the link $l$, and $r_{p}=1/-1$ for the absence/occurrence of a syndrome error of the $Z$ stabilizer at the plaquette $p$.

   In terms of the stat-mech model, the bit-flip error threshold of the toric code with $O(d)$ rounds of syndrome extraction is mapped to the confinement transition of the R4bIM at $d\to \infty$ with increasing $p$ and $q$. At low error rates, the stat-mech model is in the deconfined (Higgs) phase, signaled by the perimeter-law decay of the Wilson loop order parameter $\langle W(\gamma)\rangle\equiv\langle\prod_{l\in\gamma}\sigma_l\rangle \sim e^{-A|\gamma|}$, where $\gamma$ is a closed loop in the cubic lattice, $|\gamma|$ is its length, and $A>0$ is some constant. Here, $\langle\dots\rangle$ represents both averaging over the partition function $Z_{3D}(E)$ and then averaging over the disorder $E$. At high error rates, the model enters the confined phase where the expectation value of the Wilson loop $W(\gamma)$ decays with the area $\mathfrak{S}_\gamma$ that the loop $\gamma$ encircles: $\langle W(\gamma)\rangle\sim e^{-B\mathfrak{S}_\gamma}$, where $B>0$ is some constant. The critical point of the confinement transition is around $p^*\approx 0.033$ in the case of $p=q$~\cite{ohno2004NuclearPhysicsBb}.

\section{Toric codes under tCNOT gate with perfect syndromes}\label{sec:statmech}

   We now generalize the stat-mech mapping to logical circuits involving the tCNOT gate between two toric code blocks.  Physically, the tCNOT gate is implemented via physical CNOT gates between every pair of physical qubits (see Fig.~\ref{fig:problem_statement}(a)). Since the bit-flip operation on the control qubit does not commute with the CNOT gate, physical errors spread from the control to the target block. We consider a specific bit-flip error model that manifests this error-spreading mechanism, and study its effect on the error threshold.
The logical error rate still takes the form of Eq.~\eqref{eq:logicalprob}. However, we now need to determine the form of the cycle $C$ in the presence of logical gates, in order to sum over all cycles. The key difference from the memory case is that the notion of syndrome must be generalized to incorporate the dynamical propagation of errors through the transversal gate. We therefore first introduce the bit-flip error model of interest and then review error syndromes for a tCNOT gate.

\subsection{Persistent bit-flip  error model}\label{sec:persist}

   For quantum memories, bit-flip errors between two consecutive rounds of syndrome extractions are modeled via the following quantum channel:
\begin{equation}\label{eq:bitflip}
    \mathcal{N}_{p}\equiv\circ_i \mathcal{N}_{p,i},\ \mathcal{N}_{p,i}(\rho)\equiv(1-p)\rho+p X_i\rho X_i.
\end{equation}
Here, $p$ is the cumulative probability that a physical bit-flip error occurs between two consecutive rounds of syndrome extraction, and $i$ labels different physical qubits in the system. However, since bit-flip operation on the control qubit does not commute with the physical CNOT gate, the error model in a logical circuit needs to further specify the temporal location of the bit-flip errors.

   Therefore, we consider an error model that captures the spread of errors, in which bit-flip errors occur both before and after the tCNOT gate. To directly compare with the case of quantum memory, we evenly split the error rate of channel $\mathcal{N}_{p}$ into two bit-flip channels $\mathcal{N}_{\tilde{p}}$ before and after the tCNOT gate:
\begin{equation}
    \mathcal{N}_{\tilde{p}}\circ \mathcal{N}_{\tilde{p}}=\mathcal{N}_{p},
\end{equation}
as illustrated in Fig.~\ref{fig:problem_statement}(b).
In order for the combined error channels to have a net bit-flip probability $p$, we must have 
\begin{equation}\label{eq:pxpxbar}
    p=2\tilde{p}(1-\tilde{p}).
\end{equation}
We will subsequently refer to such an error model as the \textit{persistent bit-flip error}\footnote{When referring to the bit-flip error without the word  ``persistent", we assume the bit-flip channel in Eq.~\eqref{eq:bitflip} occurs only after the tCNOT gate, which is the case in the noise model we study in Sec.~\ref{sec:noisysyn}. }. 

To make a direct comparison with the memory thresholds, we assume that every physical CNOT gate in the tCNOT gate is free from noise. In reality, the errors of the physical CNOT gates can be incorporated into the error channels $\mathcal{N}_{\tilde{p}}$ before and after them.

\subsection{Detectors of toric codes across a tCNOT gate}\label{sec:errordetect}

The main challenge in applying stat-mech mapping to logical circuits lies in the need to track dynamical error propagation through gates. To this end, we import the quantum circuit simulation notion of a detector~\cite{gidney2021Quantuma} and integrate it into the stat-mech mapping procedure. In our case, a detector associated with a stabilizer measurement outcome is a product of that outcome with earlier stabilizer measurement outcomes that back-tracks the flow of the measured Pauli operator in the circuit. 
Without errors, each detector's value is deterministically 1. When an error $E$ occurs in the circuit, a detector can be triggered, causing its value to flip to -1. The collection of detector values, $\{d\}$, serves as the syndrome. Therefore, once the detectors and its relation with every error $E$ in the error model is specified, so will the cycles $C$. 

\begin{figure}[h]
    \centering
\includegraphics[width=0.7\columnwidth, page=1]{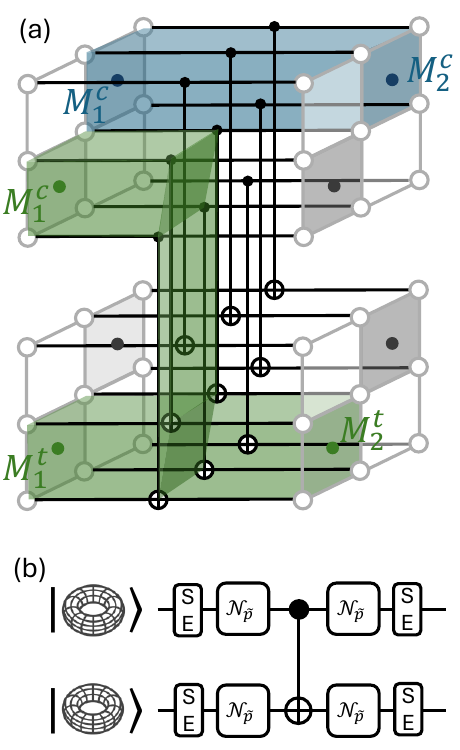}
    \caption{(a) The implementation of the tCNOT gate between two rounds of syndrome extractions. The black dots at the centers of the shaded plaquettes represent measurements of the $Z$ stabilizers defined on these plaquettes. The choices of spacetime detectors $d^c$ and $d^t$ for the control and target code blocks between the two rounds of SE are marked in blue and green, respectively. Due to the tCNOT gate, the spacetime detector of the target block, $d^t$, extends to the control block, whose value is obtained from multiplying three $Z$ stabilizer measurement outcomes $d^t=M^c_1M^t_1M^t_2$. To avoid visual overlap, we sketch the spacetime detectors $d^c$ and $d^t$ in two adjacent plaquettes that host $Z$ stabilizers. (b) The logical circuit with tCNOT gate, persistent bit-flip noise channels $\mathcal{N}_{\tilde{p}}$ and two rounds of perfect syndrome extractions (SE), during which the weight-4 $Z$ stabilizers are measured. }
    \label{fig:problem_statement}
\end{figure}

In the context of a logical circuit with a tCNOT gate, the spacetime detectors are fixed in the following way~\cite{cain2024Phys.Rev.Lett.a,sahay2025PRXQuantum,cain2025a,serra-peralta2025,turner2025}. Since we consider only bit-flip errors, we backtrack the spread of the measured $Z$ stabilizers. In this way, the spacetime detectors between the two adjacent syndrome extraction rounds are the ones shown in Fig.~\ref{fig:problem_statement}(a): the spacetime detector $d^{c}$ of the control code block is the product of the measurement outcomes $M^c_{1/2}$ of the $Z$ stabilizers before/after the tCNOT gate:
\begin{equation}\label{eq:dc}
d^c=M^c_1M^c_2. 
\end{equation}
Meanwhile, the spacetime detector $d^t$ of the target code block to be the product is now
\begin{equation}\label{eq:dt}
    d^t=M^c_1M^t_1M^t_2,
\end{equation}
where $M^t_{1,2}$ are outcomes of the $Z$ stabilizer measurements before and after the tCNOT gate. The superscripts denote the control code ($c$) and the target code ($t$), respectively (alternatively, one can also choose $ d^t=M^c_2M^t_1M^t_2$).  A visualization of the detectors is shown in Fig.~\ref{fig:problem_statement}(a). 
The spacetime detectors directly incorporate both error-propagation mechanisms arising from the tCNOT gate. Namely, a bit-flip error that occurs in the error channel $\mathcal{N}_{\tilde{p}}$ before the tCNOT gate in the control block now triggers both $d^c$ and $d^t$.

\subsection{Stat-mech mapping for perfect syndromes and persistent bit-flip errors}
\label{sec:atmodel}

\begin{figure}
    \centering
    \includegraphics[width=\linewidth, page=3]{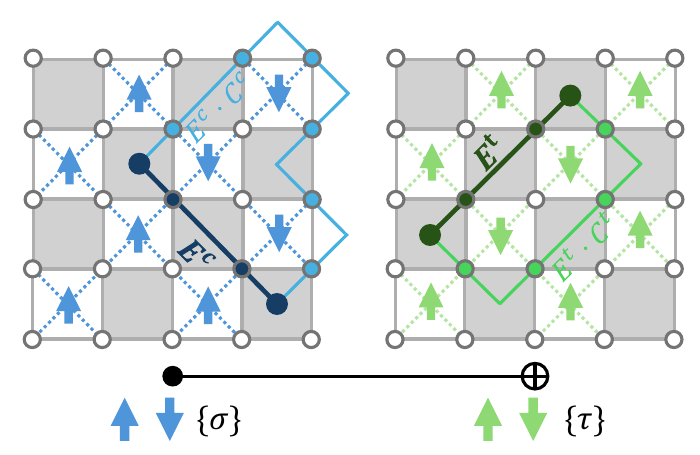}
    \caption{Stat-mech mapping of two toric code blocks undergoing a tCNOT gate, where the probability distribution of $E^c$ and $E^t$ is correlated due to the tCNOT gate. Here, the physical qubits are denoted by gray circles living on the lattice marked by solid lines. Bit-flip errors $E^{c/t}$ and $E^{c/t}\cdot C^{c/t}$ of the physical qubits in the control/target code blocks are marked by gray circles with colored fillings. Syndromes triggered by $E^{c/t}$ are marked by dots at the center of shaded plaquettes, which host $Z$ stabilizers. Trivial cycles $C^{c/t}$ in control/target blocks are parameterized by two sets of Ising spins $\{\sigma\}$ and $\{\tau\}$, which are located at lattice sites of the dotted-line lattice, and are denoted by arrows in blue/green. }
    \label{fig:statmech2}
\end{figure}

With the detectors and error model fixed, we now construct the stat-mech model of the two toric code blocks across the tCNOT gate. We denote the bit-flip errors that occur during the four error channels $\mathcal{N}_{\tilde{p}}$ in Fig.~\ref{fig:problem_statement}(a) as $E^{c,t}_{1,2}$, where $c/t$ indicates the control/target code blocks and $1/2$ denotes the error before/after the tCNOT gate. Given the form of the spacetime detectors $d^{c/t}$, the errors that trigger the spacetime detectors in the control and target blocks are 
\begin{equation}\label{eq:ecet}
E^c= E^c_1\cdot E^c_2 \text{ and } E^t= E^t_1\cdot E^t_2\cdot E^c_1,
\end{equation}
respectively. Since every physical bit-flip error occurs independently, the probability $\prob(E^c,E^t)$ is a product of independent probability distributions between pairs of physical qubits in the two code blocks that undergo physical CNOT gates:
\begin{equation}
    \prob(E^c,E^t)=\prod_l p(s^c_l,s^t_l),
\end{equation}
where $l$ is the index of the spatial location of the physical qubits, which now on the links of a rotated square lattice (marked in dotted lines in Fig. \ref{fig:statmech2}), $s_{1,2}^{c,t}=1/-1$ to denote the absence/occurrence of a bit-flip in each respective error channels, and $p(s^c_l,s^t_l)$ denotes the joint probability distribution of bit-flip errors occurring to the two physical qubits at $l$. 

We focus on a pair of physical qubits in the control and target blocks. Since the error channels $\mathcal{N}_{\tilde{p}}$ are identical, the errors in the two channels follow the same probability distribution:
\begin{equation}\label{eq:ps}
    p(s_{1,2}^{c,t})=\frac{1+(1-2\tilde{p} )s^{c,t}_{1,2}}{2}.
\end{equation}
Denoting the cases where the combined errors will not trigger/trigger the control and target spacetime detectors as $s^{c,t}=1/-1$, we have
\begin{align}\label{eq:pscst}
    p(s^c,s^t)=&\sum_{s_{1,2}^{c,t}=\pm 1}p(s_1^c)p(s_2^c)p(s_1^t)p(s_2^t)\nonumber\\
    &\quad\times\delta(s^c-s_1^cs_2^c)\delta(s^t-s^t_1s^t_2s^c_1)\nonumber\\
    =& \frac{1+(1-2\tilde{p})^2s^c+(1-2\tilde{p})^3s^t(1+s^c)}{4}.
\end{align}
In the first equality, the delta functions enforce the two conditions in Eq.~\eqref{eq:ecet}, while the second equality carries over the summation using Eq.~\eqref{eq:ps}. The joint distribution cannot be factored into a product of individual probability distributions, so $s^c$ and $s^t$ are correlated random variables. This is exactly due to the shared error $E^c_1$ between $E^c$ and $E^t$ in Eq.~\eqref{eq:ecet}. 

\begin{figure}
    \centering
    \includegraphics[width=\linewidth]{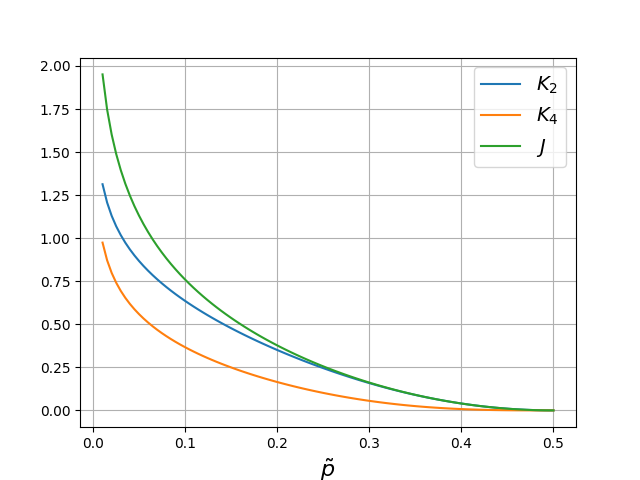}
    \caption{Coupling constants $K_2$ and $K_4$ in the random AT model Hamiltonian in Eq.~\eqref{eq:atmodel}, whose relations with $\tilde{p}$ are given in the Appendix Eq.\eqref{eq:abc}. Coupling of RBIM, $J=\frac{1}{2}\ln\frac{1-2\tilde{p}(1-\tilde{p})}{2\tilde{p}(1-\tilde{p})}$, is shown as a reference. }
    \label{fig:k2k4plot}
\end{figure}

With the representation of errors and the spacetime detectors $d^{c,t}$, we can now find a parameterization of the cycle $C$. With perfect syndromes, the spacetime detector on the plaquette $p$ in the control/target code block is triggered depending on the sign of $d^{c/t}_p=\prod_{l\in p} s^{c/t}_{l}$. Therefore, the following transformation of the physical errors will preserve the value of $d^{c/t}_p$:
\begin{equation}\label{eq:gaugetrans}
    s^c_l\to s^c_l\sigma_{l_1}\sigma_{l_2},\ s^t_l\to s^t_l\tau_{l_1}\tau_{l_2},
\end{equation}
where $\{\sigma=\pm1\},\{\tau=\pm1\}$ are two sets of Ising spins that live on the lattice sites, and $l_1,l_2$ are the two lattice sites at the two ends of the link $l$, similar to the labeling in Fig.~\ref{fig:statmech}(b). In this way, the trivial cycle $C$ can be separated into two parts: $C^c$ for the control block parameterized by $\{\sigma\}$, and  $C^t$ for the target code block parameterized by $\{\tau\}$, see Fig.~\ref{fig:statmech2}.
The stat-mech model can then be obtained via rewriting the summation of cycles using the Ising spins:
\begin{align}
    &\sum_{C^c\in S^c, C^t\in S^t}\prob(E^c\cdot C^c, E^t\cdot C^t)\nonumber\\
    &=\sum_{\{\sigma\},\{\tau\}}\prod_l p(s^c_l\sigma_{l_1}\sigma_{l_2},s^t_l\tau_{l_1}\tau_{l_2})\nonumber\\
    &\propto Z(E^c,E^t)=\sum_{\{\sigma\},\{\tau\}}e^{-H({\{\sigma\},\{\tau\}}|E^c,E^t)},
\end{align}
where $S^{c,t}$ denotes the stabilizers in the control and target code blocks, respectively. The disordered Hamiltonian is a 2D Ashkin-Teller (AT) model on the square lattice with random link couplings~\cite{ashkin1943Phys.Rev.,nobre1993J.Phys.A:Math.Gen.,wiseman1993Phys.Rev.E,wiseman1995Phys.Rev.E,bellafard2012Phys.Rev.Lett.,zhu2015Phys.Rev.B}:
\begin{equation}
    \begin{aligned}
        &H({\{\sigma\},\{\tau\}}|E^c,E^t)= -K_2\sum_l s^c_l\sigma_{l_1}\sigma_{l_2} \\
        &\quad\quad -K_4\sum_ls^t_l\tau_{l_1}\tau_{l_2} -K_4\sum_ls^t_ls^c_l\tau_{l_1}\tau_{l_2}\sigma_{l_1}\sigma_{l_2},
    \end{aligned}
    \label{eq:atmodel}
\end{equation}
where the coupling constants $K_2$ and $K_4$ are functions of $\tilde{p}$ (see Appendix~\ref{appendix:coupling_constants} for their explicit forms), and $s^{c,t}_l$ are quenched random variables following the probability distribution in Eq.~\eqref{eq:pscst}. 
We note that random AT model also arises from stat-mech mapping of toric code quantum memory under depolarizing noise~\cite{bombin2012Phys.Rev.X} and both coherent and incoherent bit-flip noises~\cite{chen2024PhysRevB,behrends2025}. Our model in Eq.~\eqref{eq:atmodel} differs from the existing models in the coupling constants and disorder distributions.
Compared to RBIM for the memory case in Eq.~\eqref{eq:sumc}, the two coupling constants in the random AT model satisfy
\begin{equation}\label{eq:k4k2relation}
    K_4<K_2<J=\frac{1}{2}\ln\frac{1-2\tilde{p}(1-\tilde{p})}{2\tilde{p}(1-\tilde{p})}
\end{equation}
for $\tilde{p}$ around the memory threshold (see Fig.~\ref{fig:k2k4plot}). The relations between these coupling strengths have direct implications for the error thresholds of the control and target code blocks, as shown in the subsequent discussion.

The stat-mech model allows us to predict the error thresholds of the two code blocks, even before performing the numerical simulation.
In light of Eq.~\eqref{eq:logicalerror}, the logical error rate for the control/target blocks now maps to the sum over $\overline{e^{-F_{DW}^{\sigma/\tau}}}$, where $F^{\sigma/\tau}_{DW}$ is the domain wall free energy of a global domain wall of the spins $\sigma/\tau$, and the disorder average is taken over both $E^c$ and $E^t$ (i.e. $s^c$ and $s^t$). Therefore, the error thresholds of the control and target code blocks, $p^c_\text{c}$ and $p^t_\text{c}$,  correspond to the order-disorder phase transition of the $\sigma$ and $\tau$ spins, respectively. 
From Eq.~\eqref{eq:k4k2relation}, we expect that the target block will first reach the error threshold with increasing $\tilde{p}$. Since the critical point is within the ordered phase of $\{\sigma\}$ spins, we expect the coupling term between the $\{\sigma\}$ and $\{\tau\}$ spins to enhance the error threshold of the target block $p^t_\text{c}$.
On the other hand, since $K_2<J$, the error threshold of the control layer $p^c_\text{c}$ should also be reduced compared to the case of quantum memory, $p_\text{th}\approx 0.109$.

\subsection{Numerical investigations of the error thresholds}\label{sec:numerics}
\subsubsection{Monte Carlo simulation of the disordered AT model}

With the stat-mech model being constructed, we now conduct a numerical study to quantify the error thresholds of the two code blocks. 
Specifically, we perform a Monte Carlo (MC) simulation of the disordered AT model in Eq.~\eqref{eq:atmodel}, with disorders sampled from the distribution shown in Eq.~\eqref{eq:pscst}. 
Unlike the conventional AT model, where $K_2$ and $K_4$ are independent parameters, our model depends on a single parameter $p$: $K_2$ and $K_4$ are both functions of $p$ (see Fig.~\ref{fig:k2k4plot}).  At each $p$, we first sample disorder configurations on all bonds $\{s_l^t, s_l^c\}$ following Eq.~\eqref{eq:pscst}. We then to equilibrate and sample the spin fields $\tau, \sigma$ with Hamiltonian described in Eq.~\eqref{eq:atmodel} given quenched disorders $\{s_l^t, s_l^c\}$. 
To ensure the MC simulation equilibrates, we introduce a hypothetical temperature $\beta$. The target distribution that we want to sample from is $e^{-\beta H}$ with $\beta=1$.  We tune $\beta$ to anneal the system from a high temperature to the target temperature $\beta=1$. We found this additional annealing equilibration stage to be crucial for reaching equilibrium.  
  
\begin{figure}
    \centering
\includegraphics[width=0.9\columnwidth, page=5]{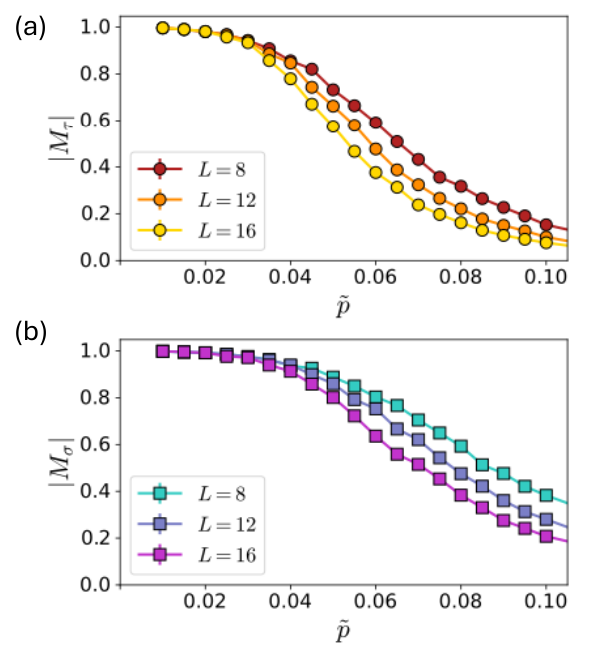}
    \caption{ Magnetization of (a) $\tau$ and (b) $\sigma$ ($|M_\tau|$ and $|M_\sigma|$)  versus persistent noise strength $\tilde{p}$ for system sizes $L=8, 12, 16$.}
    \label{fig:magnetization}
\end{figure}

To observe the order-to-disorder transition, we analyze the magnetization of the spin system as a function of $p$. As shown in Fig.~\ref{fig:magnetization}, the magnetization $\tau$ and $\sigma$ spin fields $|M|$ both saturate to $1$ in the low noise limit ($p=0$). Since we start the MC from a random initialization, $|M|=1$ in the low-$p$ limit confirms the convergence of the simulation.  As the noise level increases, $\tau$ magnetization, corresponding to the target code block, first starts decreasing and eventually vanishes. The magnetization of $\sigma$, corresponding to the control code block, tolerates a larger noise level before it starts to deviate from $|M|=1$. This observation is consistent with the fact that bit-flip errors propagate from the control code block to the target code block through the tCNOT operation, resulting in a lower threshold for the target code than for the control code.

The major benefit of stat-mech mapping is the ability to leverage the established practice of finite-size scaling (FSS) to determine error thresholds in the infinite-code-distance limit. To obtain these thresholds, we need to collapse the magnetization of different system sizes $L=8, 12, 16$ into a unique scaling form~\cite{landau2021}:
\begin{equation}
\label{eq:collapse}
    ML^{\beta/\nu} =  \mathcal{F}(\epsilon L^{1/\nu}), 
\end{equation}
where $\epsilon=\frac{\tilde{p}-\tilde{p}_\text{c}}{\tilde{p}_\text{c}}$ is the rescaled distance from the critical point and $\mathcal{F}$ is some universal scaling function. We optimize three parameters $\beta, \nu, \tilde{p}^\text{c}$ to infer the critical error rate $p$ in the thermodynamic limit (infinite code distance).

\begin{figure}
    \centering
\includegraphics[width=0.8\columnwidth, page=6]{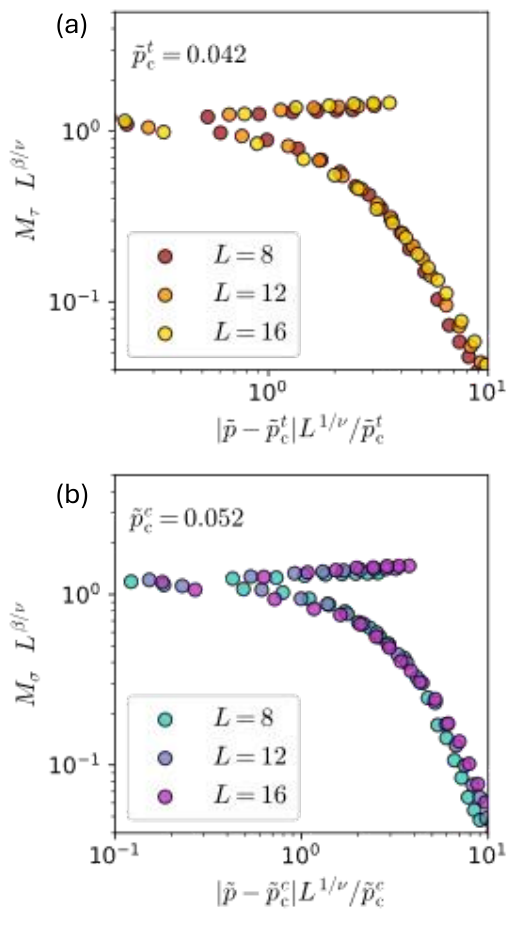}
    \caption{Finite size collapse of magnetization for (a) $\tau$ and (b) $\sigma$ according to scaling formula Eq.~\ref{eq:collapse}. The collapse is obtained from the following choice of parameters: $\beta=0.252, \nu=1.8, \tilde{p}^t_\text{c}=0.042,$ and $\tilde{p}^c_\text{c}=0.052$. } 
    \label{fig:collapse}
\end{figure}

As shown in Fig~\ref{fig:collapse}, we observe that the rescaled magnetization data achieve a remarkable collapse to universal functional forms around $\tilde{p}_\text{c}^c=0.052$ (or $p^c_\text{c}=0.099$, see Fig.~\ref{fig:collapse}(a)) and $\tilde{p}^t_\text{c}=0.042$ (or $p^t_\text{c}=0.080$, see Fig.~\ref{fig:collapse}(b)). These are compelling evidence of universal behavior in the vicinity of these critical points. Therefore, these critical points can be interpreted as upper bounds on the decoding thresholds for control ($\sigma$) and target ($\tau$) code blocks in the large-code-distance limit. 
Compared to the threshold of quantum memory, $p_\text{th}=0.109$, the thresholds of the control and the target blocks are lowered by $\frac{0.109-0.099}{0.109}=9.2\%$ and $\frac{0.109-0.080}{0.109}=26.6\%$, respectively.
These results confirm our theoretical expectation at the end of Sec.~\ref{sec:atmodel} regarding the hierarchy of the reduced thresholds. At the same time, our results establish that tCNOT gates do not lower the thresholds to a devastating degree.

Our finite-size scaling results also highlight the benefit of correlated decoding. If one decodes the target block separately, the detectors of the target block effectively experience three consecutive bit-flip channels of $\mathcal{N}_{\tilde{p}}$. The error threshold of such a channel is the solution to
\begin{equation}
    \tilde{p}^3+3(1-\tilde{p})^2\tilde{p}=p_\text{th}=0.109,
\end{equation}
which is $\tilde{p} = 0.039$ or $p = 0.076$.  We note that such a threshold is crudely estimated in Ref.~\cite{landahl2011} to be $\frac{2}{3}p_\text{th}=0.073$. Hence, the improvement of the error threshold due to joint decoding is $\frac{p^t_c - 0.076}{0.076} = 5.3\%$. 
In the stat-mech model, this improvement stems from the coupling term in the random AT model: since the $\{\tau\}$ spins’ critical point lies within the ordered phase of $\{\sigma\}$ spins, the coupling in Eq.~\eqref{eq:atmodel} induces an extra Ising coupling between the $\{\tau\}$ spins.

\begin{figure}
    \centering
\includegraphics[width=0.9\columnwidth, page=7]{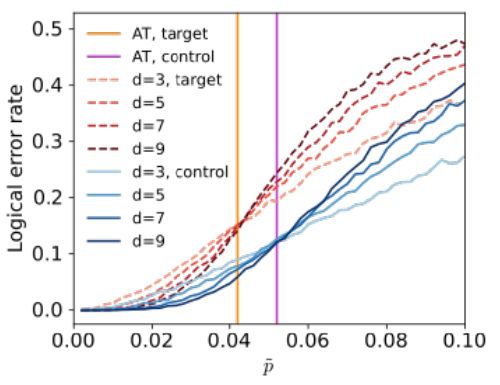}
    \caption{Logical error rate of a surface code under the same noise model using the most likely error decoder. The crossing point of curves for different system sizes indicates a decoding threshold. The gray lines represent the phase transition points predicted by Monte Carlo simulation of the random AT model. }
    \label{fig:circuit_simulation}
\end{figure}

\subsubsection{Direct simulation of error correction protocol}
   
We cross-check the thresholds predicted by the MC simulation of the AT model against those observed in a practical error-correction protocol. 
Since the error thresholds for a planar surface code and the toric code are expected to be the same~\cite{dennis2002J.Math.Phys.a}, we opt to simulate a rotated surface code for convenience. We use \texttt{stim} package~\cite{gidney2021Quantuma} to simulate and sample syndromes from the same logical circuit shown in Fig.~\ref{fig:problem_statement}(a) under the noise model described in Fig.~\ref{fig:problem_statement}(b). 
We then decode the sampled syndromes using the most likely error (MLE) decoder~\cite{cain2024Phys.Rev.Lett.a} and record the resulting logical error rates (LERs). Although the MLE decoder is not scalable and thus not suitable for practical QEC, it remains a valuable tool for validating our theoretical predictions and identifying performance bounds. 
As shown in Fig.~\ref{fig:circuit_simulation}, the LERs of the control and target code with different code distances each cross at one particular error rate, which is the observed decoding threshold. The crossings of LER for both control and target codes match the critical points obtained from FSS analysis of magnetization data from MC simulations of a stat-mech-mapped disordered AT model. Below the decoding threshold, increasing code distance suppresses logical errors.  

\section{Toric codes under tCNOT gate with syndrome errors}\label{sec:noisysyn}

\begin{figure}
    \centering
    \includegraphics[width=0.8\linewidth, page=10]{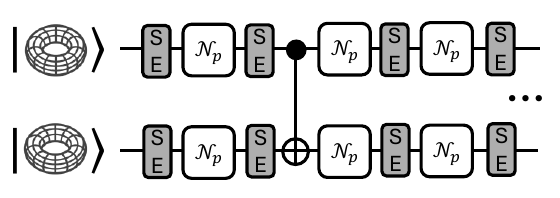}
    \caption{A toric code logical circuit with tCNOT gate, repeated noisy SE (marked by shaded boxes) and bit-flip noise channel $\mathcal{N}_{p}$.}
    \label{fig:noisycircuit}
\end{figure}

To model syndrome errors, we assume that each stabilizer measurement has a probability $q$ of being incorrect. In addition, we assume that physical bit-flip errors occur with probability $p$ only between the tCNOT gate and the next syndrome extraction (but not between the syndrome extraction and the tCNOT gate), see Fig.~\ref{fig:noisycircuit}. This error model focuses on the propagation of syndrome errors by the tCNOT gates.

\begin{figure}
    \centering
    \includegraphics[width=0.9\linewidth, page=8]{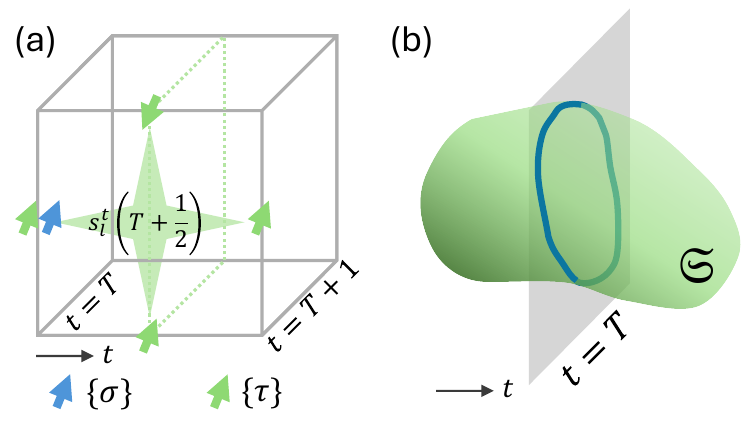}
    \caption{(a) The modified 5-body interaction term in the Hamiltonian in Eq.~\eqref{eq:coupledterms} for the $\{\tau\}$ spins due to the tCNOT gate at $t=T$.(b) Closed surfaces $\mathfrak{S}$ in the high temperature expansion of the terms involving the $\{\tau\}$ spins in the partition function $Z_{3D}(E^c,E^t)$. On the intersection between $\mathfrak{S}$ and the $t=T$ plane, the coupled terms between $\{\sigma\}$ and $\{\tau\}$ spins in Eq.~\eqref{eq:coupledterms} gives rise to a Wilson loop of $\{\sigma\}$ spins, marked by the blue loop.}
    \label{fig:3dfig}
\end{figure}

   We now construct the stat-mech model for repeated syndrome error extractions around the tCNOT gate. 
Using the expression of the detectors in Eqs.~\eqref{eq:dc} and \eqref{eq:dt}, the cycle $C$ can be parameterized using two sets of Ising spins, denoted by $\{\sigma\}$ and $\{\tau\}$, which live on the links of a 3D cubic lattice.
Summing over all possible cycles, we arrive at the partition function of the following stat-mech model (see Appendix~\ref{appendix:noisy_syndrome} for derivation):
\begin{align}\label{eq:planedef}
    &Z_{3D}(E^c,E^t)=\sum_C\prob((E^c,E^t)\cdot C)\nonumber\\
    &\quad\quad\quad\quad\quad\ =\sum_{\{\sigma\},\{\tau\}}e^{-H_{3D}(\{\sigma\},\{\tau\}|E^c,E^t)},\nonumber\\
    &H_{3D}(\{\sigma\},\{\tau\}|E^c,E^t)=H_{3D}(\{\sigma\}|E^c)\nonumber\\
    &\quad\quad\quad +H_{3D}^{t\leq T}(\{\tau\}|E^t)+H_{3D}^{t> T}(\{\sigma\tau\}|E^t).
\end{align}
Here, each individual Hamiltonian on the RHS is the one of the R4bIM given in Eq.~\eqref{eq:R4bIM}, while $t\geq T$ in $H_{3D}^{t\leq T}$ means that this Hamiltonian only contains terms that are defined on the plaquettes at or before $t=T$, and similarly for $H_{3D}^{t>T}$. In fact, the Hamiltonian $H_{3D}(\{\sigma\},\{\tau\}|E^c,E^t)$ represents two decoupled R4bIMs with a spin permutation defect plane at $t=T$ that exchanges $\tau\lrar \tau\sigma$. Redefining $\tau$ to be $\tau\sigma$ at $t>T$ in the partition function further decouples the two flavors of spins there. As a result, they are only coupled via the following set of 5-body interaction terms between $t=T$ and $t=T+1$ (see Fig.~\ref{fig:3dfig}(a)):
\begin{align}\label{eq:coupledterms}
    -J\sum_{l}&s^t_{l}(T+\frac{1}{2})\sigma_l(T)\times\nonumber\\
    &\tau_l(T)\tau_l(T+1)\tau_{l_1}(T+\frac{1}{2})\tau_{l_2}(T+\frac{1}{2}).
\end{align}

   Due to the coupling terms in Eq.~\eqref{eq:coupledterms}, we expect the error threshold of the target to be modified near $t=T$. However, obtaining a direct analytic statement of the error threshold is difficult, and a numerical simulation of the entire stat-mech model in Eq.~\eqref{eq:planedef} is required. Nevertheless, techniques in statistical mechanics, such as high- and low-temperature expansions, often shed light on critical points by connecting seemingly different stat-mech models~\cite{savit1980Rev.Mod.Phys.}. This includes the renowned Kramers-Wannier duality in the 2D Ising model, where the critical point can be computed exactly by comparing the high- and low-temperature expansions. 

   Following these inspirations, we perform a high-temperature expansion of the terms involving the $\{\tau\}$ spins in the Hamiltonian, with the assumption that the $\{\sigma\}$ spins are in the deconfined (decodable) phase when the $\{\tau\}$ spins reach the critical point.
In this way, the partition function in Eq.~\eqref{eq:planedef} can be written as
\begin{align}\label{eq:hightempexp}
    &Z_{3D}(E^c,E^t)\propto \nonumber\\
    &Z_c(E^c)\sum_{\mathfrak{S}} \left[4\tanh (J,K)\right]^{|\mathfrak{S}|}\prod_{p\in\mathfrak{S}}(s^t_p,r^t_p)\overline{\prod_{l\in \mathfrak{S}\cap (t=T)}\sigma_l}.
\end{align}
Here, the summation above is over the closed surface $\mathfrak{S}$ on the 3D cubic lattice. $\tanh^{|\mathfrak{S}|} (J,K)$ is a shorthand notation that represents the product of $|\mathfrak{S}|$ factors of  $\tanh J$ and $\tanh K$. The factor $4$ arises because each classical spin is shared by two plaquettes; summing over them yields a factor of $2^2=4$. Similarly, $\prod_{p\in\mathfrak{S}}(s^t_p,r^t_p)$ represents the product of the random signs of $s^t$ and $r^t$ in the Hamiltonian terms. $Z_c(E^c)=\sum_{\{\sigma\}}e^{-H_{3D}(\{\sigma\}|E^c)}$ is the partition function of a single R4bIM, and $\overline{\cdots}$ denote the thermally averaged value with respect to $H_{3D}(\{\sigma\}|E^c)$.
$\mathfrak{S}\cap (t=T)$ denotes the loop at the intersection between the surface $\mathfrak{S}$ and the time slice $t=T$, see Fig.~\ref{fig:3dfig}(b). Since the $\{\sigma\}$ spins are in the deconfined phase, we expect that the thermal average for a typical $E^c$ follows the perimeter-law, so that $\overline{\prod_{l\in \gamma }\sigma_l}\sim e^{-A|\gamma|}$ for any loop $\gamma$ with some constant $A>0$. Therefore, when the $\{\sigma\}$ spins are summed over, the $\{\tau\}$ spins are in an effective R4bIM, whose coupling constant $J$ is modified as 
\begin{equation}
    \tanh J\to \tanh J_\text{eff}= e^{-A}\tanh J 
\end{equation}
for all the terms in Eq.~\eqref{eq:coupledterms} that are located near the plane defect. Intuitively, this effect can be regarded as an increase in local temperature near $t=T$. Hence, with increasing error rates $p$ and $q$, the $\{\tau\}$ spins near $t=T$ will enter a local confined phase before the confinement transition for the rest of the system.
From the decoding perspective, this implies that the target code block has a finite but lower error threshold near $t=T$ than the rest of the system.

For a rough, conservative estimate for the local error threshold of the target block near $t=T$, we consider the situation where the Wilson loop of the $\{\sigma\}$ spin renormalizes the coupling constant $J$ to $J_\text{eff}$ in a considerable region around $t=T$. If we limit ourselves to the case where the syndrome error rate $q$ equals the bit-flip rate $p$, setting 
$J=K$, the target block reaches threshold when $J_\text{eff}(p)$ reaches its critical value:
\begin{equation}\label{eq:pct}
    \tanh J_\text{eff}(p^t_\text{c})=e^{-A(p^t_\text{c})}(1-2p^t_\text{c})=\tanh J(p^*)=1-2 p^*.
\end{equation}
This implies that the reduction of error threshold from that of memory with imperfect syndromes, $p^*=0.033$ will be moderate as long as the loop tension $A$ is small. 
The numerical study of R4bIM in Ref.~\cite{ohno2004NuclearPhysicsBb} found that $A$ is of order $0.01$ when $p$ is close but smaller than $p^*$. Therefore, we estimate the solution to Eq.~\eqref{eq:pct} by setting $A=0.01$, which yields $p_\text{c}^t=0.028$. Hence, even under conservative assumptions, the decrease in the target block's error threshold due to the spread of syndrome errors is not significant.
We leave the direct numerical simulation of the local error threshold for future work.

\section{Conclusion and outlook}\label{sec:conclusion}

\begin{figure}
    \centering
    \includegraphics[width=.9\linewidth, page=9]{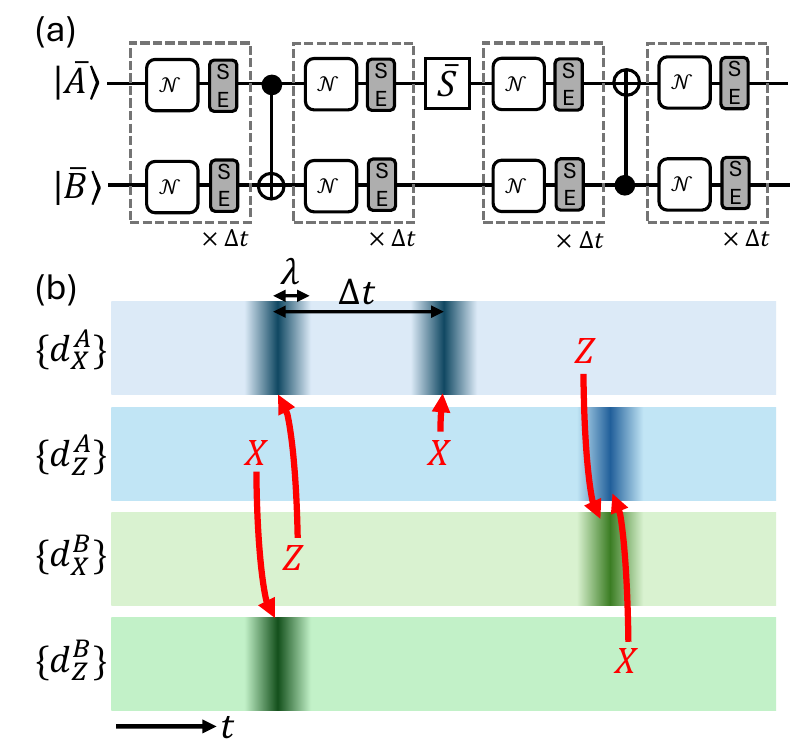}
    \caption{(a) A noisy transversal Clifford logical circuit between two code blocks $A$ and $B$ consisting of two tCNOT gates and a (fold-)transversal $S$ gate. Every pair of transversal gates is separated by $\Delta t$ rounds of syndrome error extractions and generic physical error channels $\mathcal{N}$. (b) Sketch of the impact of transversal gates on the stat-mech model that is mapped from the circuit in (a). The $X/Z$ spacetime detectors of the two code blocks $A$ and $B$, denoted by $\{d^{A/B}_{X/Z}\}$, give rise to 4 different sets of classical spins in total (see Appendix~\ref{app:foldtransversal} for details), which are coupled at the plane defects induced by the logical operations. The propagation of Pauli errors by the transversal gates is marked by red arrows. The impact region $|t-T|
    <\lambda$ due to increased effective local temperature is marked by darker colors for each plot. }
    \label{fig:timeline}
\end{figure}

   In this work, we have generalized the statistical-mechanical mapping, a foundational tool for establishing rigorous, decoder-independent error thresholds for quantum memories, to logical circuits with transversal gates. Our work provides the first extension of stat-mech mapping to circuits that perform logical operations. The key insight underlying this generalization is that each transversal gate introduces a permutation defect of the classical spins at the time slice where the gate is applied, modifying the stat-mech model only locally in time.
   
   We demonstrated this framework by deriving stat-mech models for two toric code blocks undergoing a tCNOT gate under two complementary error models. For persistent bit-flip errors with perfect syndromes, the decoding problem maps to a 2D random AT model, and Monte Carlo simulations with FSS yield error thresholds of $p_\text{c}^c=0.099$ and $p_\text{c}^t=0.080$ for the control and target code blocks, respectively. For bit-flip errors coexisting with syndrome errors, the circuit maps to a 3D R4bIM with a plane defect, and we conservatively estimate the target block threshold at $p_\text{c}^t\geq 0.028$.
   Comparing with the corresponding memory thresholds ($p_\text{th}=0.109$ and $p^*=0.033$), these results establish that the error spreading by the tCNOT gate causes a moderate, non-devastating decrease in the decoding threshold---a reassuring conclusion for the viability of transversal-gate-based fault-tolerant computation.

   An essential feature of our framework is its systematic generalizability. In Appendix~\ref{app:foldtransversal}, we derive the stat-mech models for the fold-transversal Hadamard and $S$ gates of the toric code under generic single-qubit Pauli noise, showing that the transversal Hadamard gate induces a flavor-exchange defect ($\sigma^x \leftrightarrow \sigma^z$), while the transversal $S$ gate introduces a coupling between the $X$ and $Z$ error sectors identical in structure to the tCNOT plane defect. Together with the tCNOT gate, these results provide a complete stat-mech mapping for any transversal Clifford logical circuit of the toric code, where each consecutive gate is separated by $\Delta t$ rounds of syndrome extractions, as illustrated in Fig.~\ref{fig:timeline}(a).
   Crucially, each gate affects the stat-mech model only within a time window $|t-T|\leq \lambda$ with $\lambda\sim O(1)$, as sketched in Fig.~\ref{fig:timeline}(b). This locality ensures that the threshold of the full circuit can be estimated by simulating the stat-mech model with all the plane defects corresponding to the gates in the circuit.

   In Appendix~\ref{app:general}, we further extend the framework to arbitrary CSS codes that support transversal Clifford gates, using a binary matrix formalism. We prove that: (i) every transversal gate modifies the spacetime parity check matrix via a local modification that changes the kernel of the check matrix; (ii) the resulting stat-mech model differs from the memory model only at the time of the gate; and (iii) the effects of multiple transversal gates compose multiplicatively (Eq.~\eqref{eq:wmatmultiply}). These results establish that the stat-mech mapping is, in principle, applicable to any transversal Clifford logical circuit built from CSS codes, providing a systematic route to rigorous threshold analysis of fault-tolerant quantum computation.

   Looking ahead, our work opens several exciting directions. The stat-mech models derived here can be simulated directly for specific circuits of interest, thereby enabling threshold benchmarks for near-term fault-tolerant architectures. Furthermore, the connection between error thresholds and phase transitions in the stat-mech model suggests a potential link to topological phases of matter: the noisy logical circuit within and beyond the error threshold may be viewed as two distinct phases, and the determination of the threshold can be formulated as a phase-recognition task amenable to machine learning approaches~\cite{kim2024}. Finally, extending the stat-mech mapping beyond Clifford circuits, potentially by incorporating magic-state injection or non-Clifford transversal gates, remains an important open problem.

{\bf Acknowledgments}
YX thanks Kaavya Sahay and Mark Turner for helpful discussions on tCNOT decoders.
YX and E-AK acknowledge support by the NSF through the grant OAC-2118310. YZ and E-AK acknowledge support by the National Science Foundation (Platform for the Accelerated Realization, Analysis, and Discovery of Interface Materials (PARADIM)) under Cooperative Agreement No. DMR-2039380. This research is funded in part by the Gordon and Betty Moore Foundation’s EPiQS Initiative, Grant GBMF10436 to E-AK. JPS was funded by NSF DMR 2327094. This research was supported in part by grant NSF PHY-2309135 to the Kavli Institute for Theoretical Physics (KITP).

\bibliographystyle{quantum}
\bibliography{main}

\appendix 

\onecolumn
\section{Coupling constants in the Hamiltonian of the random Ashkin-Teller model}\label{appendix:coupling_constants}

In this appendix, we derive the coupling constants $K_2$ and $K_4$ in the random AT model in Eq.~\eqref{eq:atmodel}. To this end, we first rewrite the probability distribution of the random link distribution in Eq.~\eqref{eq:pscst} in an exponential ansatz:
\begin{equation}
    p(s^c,s^t)\propto e^{A s^c+Bs^t+C s^c s^t},
\end{equation}
where $A,B$ and $C$ are constants to be fixed from the equality. Expanding the exponential using the equality $e^{As}=\cosh A+s \sinh A$ for any binary number $s=\pm 1$, we have
\begin{align}
    &(1-2\tilde{p})^2=\frac{\sinh A \cosh B \cosh C+\cosh A \sinh B\sinh C}{\cosh A\cosh B\cosh C+\sinh A \sinh B \sinh C},\nonumber\\
    &(1-2\tilde{p})^3=\frac{\cosh A \sinh B \cosh C+\sinh A \cosh B\sinh C}{\cosh A\cosh B\cosh C+\sinh A \sinh B \sinh C}=\frac{\cosh A \cosh B \sinh C+\sinh A \sinh B\cosh C}{\cosh A\cosh B\cosh C+\sinh A \sinh B \sinh C}.
\end{align}
Solving the equations above, we have
\begin{align}\label{eq:abc}
    &A=\frac{1}{4}\log\frac{[1+(1-2\tilde{p})^2]^2-4(1-2\tilde{p})^6}{[1-(1-2\tilde{p})^2]^2}=K_2,\nonumber\\
    &B=C=\frac{1}{4}\log\frac{1+(1-2\tilde{p})^2+2(1-2\tilde{p})^3}{1+(1-2\tilde{p})^2-2(1-2\tilde{p})^3}=K_4.
\end{align}
In the final equalities of both lines, we redefined the constants to be $K_2$ and $K_4$, which match the coupling constants in the Hamiltonian in Eq.~\eqref{eq:atmodel} once we substitute $s^{c/t}_{l}$ according to Eq.~\eqref{eq:gaugetrans}. Both $K_2$ and $K_4$ decrease monotonically with increasing $\tilde{p}$ (see Fig.~\ref{fig:k2k4plot}), leading to order-disorder phase transitions of the $\{\sigma\}$ and $\{\tau\}$ spins. Since $K_2>K_4$, the $\{\tau\}$ spins disorders before $\{\sigma\}$ spins with increasing $\tilde{p}$.

\section{Derivation of the stat-mech model for tCNOT with syndrome errors} \label{appendix:noisy_syndrome}
To parameterize the error $E$ with syndrome errors, we use a binary number $r^{c,t}_p(t)=1/-1$ to denote that the syndrome extraction of the $Z$ stabilizer at the plaquette $p$ in the control or target block is correct/incorrect, respectively. $t\in\Z$ in the brackets denotes the time step at which syndrome extraction occurs. The probability distribution is given by $p_r(r^{c,t}_p(t))=\frac{1+(1-2q)r^{c,t}_p(t)}{2}$. The bit-flip errors are now represented by $s_l^{c,t}(t+\frac{1}{2})=\pm 1$ with probability $p_s(s_l^{c,t}(t+\frac{1}{2}))=\frac{1+(1-2p)s_l^{c,t}(t+\frac{1}{2})}{2}$. Without the tCNOT gate, the values of the spacetime detectors are given by
\begin{equation}\label{eq:stdetec}
    d^c(t+\frac{1}{2})=r^c_p(t)r^c_p(t+1)\prod_{l\in p}s_l^c(t+\frac{1}{2})\text{ and } d^t(t+\frac{1}{2})=r^t_p(t)r^t_p(t+1)\prod_{l\in p}s_l^t(t+\frac{1}{2}).
\end{equation}
For a tCNOT gate that happens right after time step $T$, the values of the spacetime detectors of the control and target blocks become
\begin{equation}\label{eq:stdetecc}
    d^c(T+\frac{1}{2})=r^c_p(T)r^c_p(T+1)\prod_{l\in p}s_l^c(T+\frac{1}{2})\text{ and } d^t(T+\frac{1}{2})=r^t_p(T)r^c_p(T)r^t_p(T+1)\prod_{l\in p}s_l^t(T+\frac{1}{2}),
\end{equation}
respectively. Note that since $r^c_p(T)$ is in three spacetime detectors, $d^c(T-\frac{1}{2})$, $d^c(T+\frac{1}{2})$ and $d^t(T+\frac{1}{2})$, the decoding graph across the tCNOT gate contains weight-3 hyperedges. Away from $t=T$, $r^c_p$ do not enter the spacetime detector $d^t(t\neq T)$ of the target block.

To obtain the stat-mech model in this case, we need to parameterize the trivial cycles $C$ that lead to the same set of syndromes $\{d^c,d^t\}$. Based on their forms in Eqs.~\eqref{eq:stdetec} and \eqref{eq:stdetecc}, we introduce two sets of Ising spins $\{\sigma\}$ and $\{\tau\}$ for every link of the cubic lattice. When $t\leq T$, the spacetime detectors are invariant under the following reparameterization via the Ising spins:
\begin{align}
    &s^{c}_l(t-\frac{1}{2})\to s^{c}_l(t-\frac{1}{2})\sigma_l(t-1)\sigma_l(t)\sigma_{l_1}(t-\frac{1}{2})\sigma_{l_2}(t-\frac{1}{2}),\ s^{t}_l(t-\frac{1}{2})\to s^{t}_l(t-\frac{1}{2})\tau_l(t-1)\tau_l(t)\tau_{l_1}(t-\frac{1}{2})\tau_{l_2}(t-\frac{1}{2}),\nonumber\\
    &r^c_p(t)\to r^c_p(t)\prod_{l\in p}\sigma_l(t),\ r^t_p(t)\to r^t_p(t)\prod_{l\in p}\tau_l(t),
\end{align}
the same as the case without the tCNOT gate.
At $t>T$, to account for $r^c_p(T)$ in the detector $d^t(T)$ in Eq.~\eqref{eq:stdetecc}, the reparameterization is changed accordingly:
\begin{align}
    &s^{c}_l(t-\frac{1}{2})\to s^{c}_l(t-\frac{1}{2})\sigma_l(t-1)\sigma_l(t)\sigma_{l_1}(t-\frac{1}{2})\sigma_{l_2}(t-\frac{1}{2}),\nonumber\\
    &s^{t}_l(t-\frac{1}{2})\to s^{t}_l(t-\frac{1}{2})\sigma_l(t-1)\tau_l(t-1)\sigma_l(t)\tau_l(t)\sigma_{l_1}(t-\frac{1}{2})\tau_{l_1}(t-\frac{1}{2})\sigma_{l_2}(t-\frac{1}{2})\tau_{l_2}(t-\frac{1}{2}),\nonumber\\
    &r^c_p(t)\to r^c_p(t)\prod_{l\in p}\sigma_l(t),\ r^t_p(t)\to r^t_p(t)\prod_{l\in p}\sigma_l(t)\tau_l(t),
\end{align}
that is, the Ising spin $\{\tau\}$ becomes $\{\tau\sigma\}$ once we go from $t\leq T$ to $t>T$. In other words, the tCNOT gate induces a domain wall that permutes the Ising spins.

Putting everything together, the probability of having the same syndrome set $\{d^c(t),d^t(t)\}$ without logical error sums up to
\begin{align}\nonumber
    &\sum_{C}\prob((E^c,E^t)\cdot C)\\
    &=\sum_{\{\sigma\},\{\tau\}}\prod_{t\neq T}\left[\prod_l p_s\left(s^c_l(t+\frac{1}{2})\sigma_l(t)\sigma_l(t+1)\sigma_{l_1}(t+\frac{1}{2})\sigma_{l_2}(t+\frac{1}{2})\right)p_s\left(s^t_l(t+\frac{1}{2})\tau_l(t)\tau_l(t+1)\tau_{l_1}(t+\frac{1}{2})\tau_{l_2}(t+\frac{1}{2})\right)\nonumber\right]\\
    &\times \prod_l p_s\left(s^c_l(T+\frac{1}{2})\sigma_l(T)\sigma_l(T+1)\sigma_{l_1}(T+\frac{1}{2})\sigma_{l_2}(T+\frac{1}{2})\right)p_s\left(s^t_l(T+\frac{1}{2})\sigma_l(T)\tau_l(T)\tau_l(T+1)\tau_{l_1}(T+\frac{1}{2})\tau_{l_2}(T+\frac{1}{2})\right)\nonumber\\
    &\times\prod_t\left[ \prod_p p_r\left(r^c_p(t)\prod_{l\in p}\sigma_l(t)\right)p_r\left(r^t_p(t)\prod_{l\in p}\tau_l(t)\right)\right].
\end{align}
Note that we have redefined the spins $\{\tau\sigma\}$ to be $\{\tau\}$ at $t>T$ in the summation. Using $p_s(s)\propto e^{Js}$ and $p_r(r)\propto e^{K r}$, where $J=\frac{1}{2}\ln\frac{1-p}{p}$ and $K=\frac{1}{2}\ln\frac{1-q}{q}$, the probability is proportional to the partition function 
\begin{align}
    &\sum_{C\in S}\prob((E^c,E^t)\cdot C)\propto Z(E^c,E^t)=\sum_{\{\sigma\},\{\tau\}}e^{-H_{3D}(\{\sigma\},\{\tau\}|E^c,E^t)}\\
    & H_{3D}(\{\sigma\},\{\tau\}|E^c,E^t)=-J\sum_{l, t}s^c_l(t+\frac{1}{2})\sigma_l(t)\sigma_l(t+1)\sigma_{l_1}(t+\frac{1}{2})\sigma_{l_2}(t+\frac{1}{2})\nonumber\\
    &\quad\quad\quad-J\sum_{l,t\neq T}s^t_l(t+\frac{1}{2})\tau_l(t)\tau_l(t+1)\tau_{l_1}(t+\frac{1}{2})\tau_{l_2}(t+\frac{1}{2})\nonumber\\
    &\quad\quad\quad -J\sum_{l}s^t_l(T+\frac{1}{2})\sigma_l(T)\tau_l(T)\tau_l(T+1)\tau_{l_1}(T+\frac{1}{2})\tau_{l_2}(T+\frac{1}{2})-K\sum_{l,t}\left[r^c_p(t)\prod_{l\in p}\sigma_l(t)+r^t_p(t)\prod_{l\in p}\tau_l(t)\right].
\end{align}
Compared to the Hamiltonian of two decoupled R4bIMs, $H_{3D}(\{\sigma\}|E^c)+H_{3D}(\{\tau\}|E^t)$, the difference in the Hamiltonian above is the terms that involve $s^t_l(T+\frac{1}{2})$, which contain the $\sigma$ spins.

\begin{figure}
    \centering
    \includegraphics[width=0.6\linewidth,page=1]{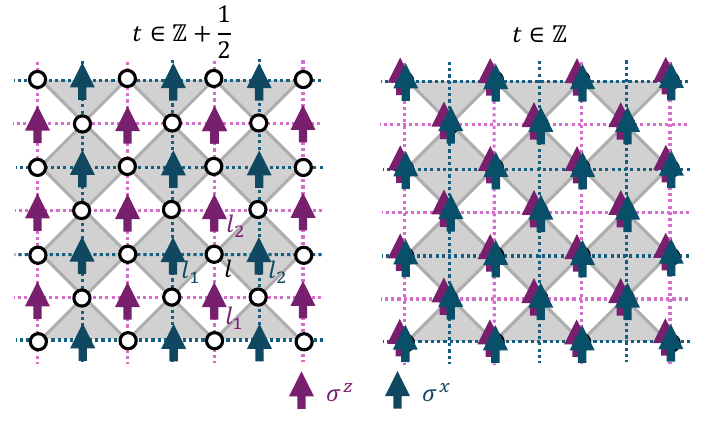}
    \caption{The stat-mech mapping of the toric code with generic Pauli and syndrome errors. Here, the physical qubits of the toric code are marked by black circles, the shaded and white squares denote weight-4 $Z$ and $X$ stabilizers, respectively. The two flavors of Ising spins, $\{\sigma^x\}$ and $\{\sigma^z\}$, each live on links of a cubic lattice. As such, at $t\in\Z+\frac{1}{2}$, the Ising spins live on two 2D square lattices that are dual to each other, and at $t\in\Z$ they overlap at the same spatial locations.}
    \label{fig:3dsm}
\end{figure}

\section{Stat-mech mapping for transversal Clifford gates and generic Pauli error channels}
\label{app:foldtransversal}
In this section, we show that our stat-mech mapping can be generalized for error thresholds of transversal Clifford logical operations under a generic single-qubit phenomenological noise model. In particular, we show that every transversal Clifford gate induces a distinct permutation defect of the Ising spins in the stat-mech model.

In general, Pauli errors between two rounds of syndrome extractions can be parameterized by the action of the quantum channel 
\begin{align}
    &\mathcal{N}\equiv \circ_i\mathcal{N}_i,\ \mathcal{N}_i:\rho\mapsto(1-p_x-p_y-p_z)\rho+p_xX_i\rho X_i +p_y Y_i\rho Y_i+p_z Z_i\rho Z_i,
\end{align}
where $i$ is the index of qubits. In this way, the bit-flip channel with probability $p$ corresponds to $(p_x,p_y,p_z)=(p,0,0)$ and the depolarizing channel with probability $p$ corresponds to $(p_x,p_y,p_z)=(p/3,p/3,p/3)$. To facilitate the stat-mech mapping, we can represent the absence/occurrence of a single-qubit Pauli error as two correlated random variables $s^x_i=\pm 1$ and $s^z_i=\pm 1$, which follows the probability distribution
\begin{align}
    p(s^x,s^z)&=\nonumber(1-p_x-p_y-p_z)\frac{1+s^x}{2}\frac{1+s^z}{2}+p_x\frac{1-s^x}{2}\frac{1+s^z}{2}+p_y\frac{1-s^x}{2}\frac{1-s^z}{2}+p_z\frac{1+s^x}{2}\frac{1-s^z}{2}\\
    &=\frac{1}{4}\left[1+(1-2p_x-2p_y)s^x+(1-2p_x-2p_z)s^xs^z+(1-2p_y-2p_z)s^z\right]\nonumber\\
    &\propto e^{J_xs^x+J_ys^xs^z+J_zs^z}.
\end{align}
In the final line, we write the probability distribution in exponential form, which will be useful in deriving the stat-mech model. The coupling constants on the exponent are related to the probabilities by
\begin{equation}
    J_x=\frac{1}{4}\ln\frac{(1-p_x-p_y-p_z)p_z}{p_xp_y},\ J_y=\frac{1}{4}\ln\frac{(1-p_x-p_y-p_z)p_y}{p_xp_z},\ J_z=\frac{1}{4}\ln\frac{(1-p_x-p_y-p_z)p_x}{p_yp_z}.
\end{equation}

When considering the persistence of the Pauli error across the transversal gate, we can evenly split the channel $\mathcal{N}$ into $\mathcal{N}=\tilde{\mathcal{N}}\circ\tilde{\mathcal{N}}$, where 
\begin{align}
    &\tilde{\mathcal{N}}\equiv \circ_i\tilde{\mathcal{N}}_i,\ \tilde{\mathcal{N}}_i:\rho\mapsto(1-\tilde{p}_x-\tilde{p}_y-\tilde{p}_z)\rho+\tilde{p}_XX_i\rho X_i +\tilde{p}_y Y_i\rho Y_i+\tilde{p}_z Z_i\rho Z_i,
\end{align}
and the probabilities $\tilde{p}_{x,y,z}$ are related to the original probabilities via
\begin{equation}
    2(1-\tilde{p}_x-\tilde{p}_y-\tilde{p}_z)\tilde{p}_\alpha+2\tilde{p}_\beta\tilde{p}_\gamma=p_\alpha,
\end{equation}
where $\alpha, \beta,\gamma$ are cyclic permutations of $x,y,z$.

To model syndrome-extraction errors, we consider an error model in which each $X$ and $Z$ stabilizer has probabilities $q_x$ and $q_z$ of being the opposite of its actual value, respectively. The absence/presence of such errors can be parameterized by parities $r^x=\pm 1$ and $ r^z=\pm 1$. They follow the probability distributions
\begin{equation}
    p(r^{x,z})=\frac{1+(1-2q_{x,z})r^{x,z}}{2}\propto e^{K_{x,z}r^{x,z}},\ K_{x,z}=\frac{1}{2}\ln\frac{1-q_{x,z}}{q_{x,z}}.
\end{equation}

Together with the physical Pauli errors, for the toric code, the parities of the detectors between two consecutive rounds of syndrome extractions read
\begin{equation}\label{eq:detect}
    d^x_p(t+\frac{1}{2})=r^x_p(t)r^x_p(t+1)\prod_{l\in p}s_l^z(t+\frac{1}{2})\text{ and } d^z_p(t+\frac{1}{2})=r^z_p(t)r^z_p(t+1)\prod_{l\in p}s_l^x(t+\frac{1}{2}).
\end{equation}
This generalizes the detectors for bit-flip errors in Eq.~\eqref{eq:stdetec}. Based on these relations, the detectors are invariant under the following reparameterization of the errors: 
\begin{align}\label{eq:reparam}
    &s^{x}_l(t-\frac{1}{2})\to s^{x}_l(t-\frac{1}{2})\sigma^x_l(t-1)\sigma^x_l(t)\sigma^x_{l_1}(t-\frac{1}{2})\sigma^x_{l_2}(t-\frac{1}{2}),\nonumber\\
    &s^{z}_l(t-\frac{1}{2})\to s^{z}_l(t-\frac{1}{2})\sigma^z_l(t-1)\sigma^z_l(t)\sigma^z_{l_1}(t-\frac{1}{2})\sigma^z_{l_2}(t-\frac{1}{2}),\nonumber\\
    &r^x_p(t)\to r^x_p(t)\prod_{l\in p}\sigma^z_l(t),\ r^z_p(t)\to r^z_p(t)\prod_{l\in p}\sigma^x_l(t).
\end{align}
Here $\{\sigma^x\}$ and $\{\sigma^z\}$ are two sets of Ising spins that live on sites of two square lattices that are dual to each other, see Fig.~\ref{fig:3dsm}. The link $l$ in $s^x_l$ and $s^z_l$ represents two links on the two square lattices that are perpendicular to each other, and $l_{1,2}$ in $\sigma^x_{l_{1,2}}$ and $\sigma^z_{l_{1,2}}$ are two sites connected by the link $l$ on the two respective lattices.

Using the standard steps of stat-mech mapping outlined in Sec.~\ref{sec:prelim}, the toric code quantum memory under the generic Pauli errors can be mapped to a stat-mech model described by the following partition function
\begin{align}\label{eq:fullstatmech}
    &Z(\{s^{x,z}\},\{r^{x,z}\})=\sum_{\{\sigma^x\},\{\sigma^z\}}e^{-H_{3D}(\{\sigma^x\},\{\sigma^z\}|\{s^{x,z}\},\{r^{x,z}\})},\nonumber\\
    &H_{3D}(\{\sigma^x\},\{\sigma^z\}|\{s^{x,z}\},\{r^{x,z}\})\equiv -J_x\sum_{l,t}s^{x}_l(t-\frac{1}{2})\sigma^x_l(t-1)\sigma^x_l(t)\sigma^x_{l_1}(t-\frac{1}{2})\sigma^x_{l_2}(t-\frac{1}{2})-(x\to z)\nonumber\\
    &-J_y\sum_{l,t}s^{x}_l(t-\frac{1}{2})s^{z}_l(t-\frac{1}{2})\sigma^x_l(t-1)\sigma^x_l(t)\sigma^x_{l_1}(t-\frac{1}{2})\sigma^x_{l_2}(t-\frac{1}{2})\sigma^z_l(t-1)\sigma^z_l(t)\sigma^z_{l_1}(t-\frac{1}{2})\sigma^z_{l_2}(t-\frac{1}{2})\nonumber\\
    &-K_x \sum_{p,t}r^x_p(t)\prod_{l\in p}\sigma^z_l(t)-K_z\sum_{p,t} r^z_p(t)\prod_{l\in p}\sigma^x_l(t).
\end{align}
The disorder distribution in the stat-mech model is the joint distribution of Pauli errors that follows $p(s^x_l(t+\frac{1}{2}),s^z_l(t+\frac{1}{2}))$ on every link $l$, and syndrome errors that follow distributions $p(r^{x}_p(t))$ and $p(r^z_p(t))$ on every plaquette:
\begin{equation}
    p(\{s^{x,z}\},\{r^{x,z}\})=\prod_t\left[\prod_lp(s^x_l(t+\frac{1}{2}),s^z_l(t+\frac{1}{2}))\prod_pp(r^{x}_p(t))p(r^{z}_p(t))\right].
\end{equation}

\begin{figure}
    \centering
    \includegraphics[width=0.5\linewidth, page=2]{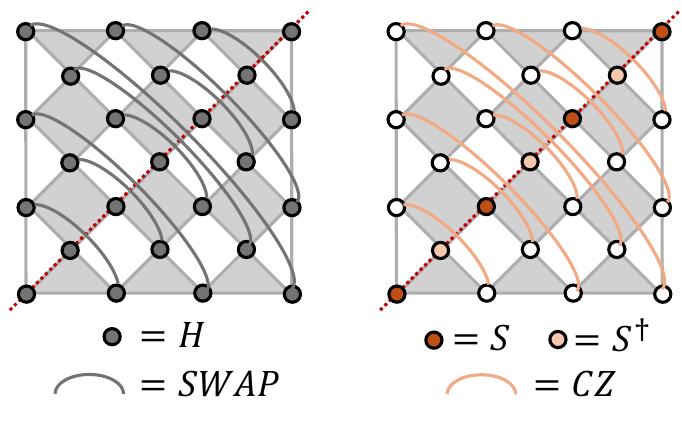}
    \caption{Fold-transversal $H$ and $S$ gates of the toric code. The reflection axis is marked by the red dotted line. }
    \label{fig:foldtgates}
\end{figure}

\subsection{Fold-transversal Hadamard gate }

On a 2D torus with equal distance in both directions, the toric code has a transversal Hadamard (t$H$) gate~\cite{moussa2016Phys.Rev.A}. Physically, it can be implemented by first applying the $H$ gate to every physical qubit, and then swapping the qubits in a reflection-symmetric way with respect to a diagonal line, see Fig.~\ref{fig:foldtgates}. For toric code on a torus with equal periodicity in both directions, the t$H$ gate simultaneously performs $H$ gates on the two encoded logical qubits.

We now derive the stat-mech model of a single toric code block under general Pauli errors with a t$H$ gate applied between $t=T$ and $t=T+1$. For simplicity, we ignore the persistence of the error, so that the Pauli error only occurs between the transversal gate and the next round of syndrome extraction at $t=T+1$, similar to the case in Fig.~\ref{fig:noisycircuit}. Due to the exchange of the $X$ and $Z$ stabilizers, the detectors across the t$H$ gate are modified as
\begin{equation}\label{eq:Hdetect}
    d^x_p(T+\frac{1}{2})=r^z_{\bar p}(T)r^x_p(T+1)\prod_{l\in p}s_l^z(T+\frac{1}{2})\text{ and } d^z_p(T+\frac{1}{2})=r^x_{\bar p}(T)r^z_p(T+1)\prod_{l\in p}s_l^x(T+\frac{1}{2}).
\end{equation}
Here $\bar p$ denotes the plaquette that is reflection-symmetric to $p$ with respect to the diagonal line. 
To preserve these detectors between $t=T$ and $t=T+1$, the reparameterization in Eq.~\eqref{eq:reparam} is modified as
\begin{align}\label{eq:Hreparam}
    &s^{x}_l(T+\frac{1}{2})\to s^{x}_l(T+\frac{1}{2})\sigma^z_{\bar l}(T)\sigma^x_l(T+1)\sigma^x_{l_1}(T+\frac{1}{2})\sigma^x_{l_2}(T+\frac{1}{2}),\nonumber\\
    &s^{z}_l(T+\frac{1}{2})\to s^{z}_l(T+\frac{1}{2})\sigma^x_{\bar l}(T)\sigma^z_l(T+1)\sigma^z_{l_1}(T+\frac{1}{2})\sigma^z_{l_2}(T+\frac{1}{2}),
\end{align}
where $\bar l$ denotes the link $l$ after the mirror-reflection with respect to the diagonal line. Compared to Eq.~\eqref{eq:reparam}, the spins $\{\sigma^x_l(T)\}$ and $\{\sigma^z_{\bar l}(T)\}$ are exchanged, and $r^{x,z}_p(t)$ remain the same.
Correspondingly, with the presence of the t$H$ gate between $t=T$ and $t=T+1$, the terms involving $s^x_l(T+\frac{1}{2})$ in the stat-mech model $H_{3D}(\{\sigma^x\},\{\sigma^z\}|\{s^{x,z}\},\{r^{x,z}\})$ are modified in the way of the RHSs in Eq.~\ref{eq:Hreparam}. In fact, the t$H$ gate induces a permutation defect of the two spin flavors across $t=T$ that exchanges $\sigma^x$ and $\sigma^z$.

Based on the stat-mech model, we now discuss the impact of the t$H$ gate on the error threshold. For simplicity, we consider a Pauli error channel where $X$ and $Z$ errors occur independently with probabilities $P_X$ and $P_Z$, in which case $(p_x,p_y,p_z)=((1-P_Z)P_X,P_XP_Z,(1-P_X)P_Z)$. Therefore, $s^x$ and $s^z$ are independent variables, and we have $J_y=0$ in the stat-mech model. Hence, the two flavors of Ising spins, $\{\sigma^x\}$ and $\{\sigma^z\}$, completely decouple except for the terms involving $s^{x,z}(T+\frac{1}{2})$ due to the t$H$ gates. We can further exchange all the spins $\sigma^x_l(t)\lrar \sigma^z_{\bar l}(t)$ for all $t\leq T$, after which the two spin flavors are completely decoupled. Since the disorder distributions of $s^{x,z}_l$ and $r^{x,z}_p$ are translation invariant, we can relabel their spatial locations $l$ and $p$ as $\bar l$ and $\bar p$ as well. Therefore, the only difference here, compared to the original stat-mech model in Eq.~\eqref{eq:fullstatmech}, is that the coupling constants $J_x$ and $J_z$, as well as $K_x$ and $K_z$, are exchanged for all the terms that involve the terms that are located within $t\leq T$. Consequently, both flavors of spins will locally confine once the error rates exceed the error threshold of quantum memory.

\subsection{Fold-transversal $S$ gate}
Additionally, the toric code has a transversal $S$ (t$S$) gate~\cite{moussa2016Phys.Rev.A}. Physically, it can be implemented by first applying $S$ or $S^\dagger$ gates to every physical qubit on a diagonal, while also applying controlled-$Z$ ($CZ$) gates between pairs of qubits that are reflection-symmetric with respect to the diagonal line, see Fig.~\ref{fig:foldtgates}. For toric code on a torus with equal periodicity in both directions, the t$S$ gate simultaneously performs $S$ gates on the two encoded logical qubits.

Now consider a t$S$ gate is applied between $t=T$ and $t=T+1$ in a single toric code block.  the detectors across the t$S$ gate are modified as
\begin{equation}\label{eq:Sdetect}
    d^x(T+\frac{1}{2})=r^x_p(T)r^z_{\bar p}(T)r^x_p(T+1)\prod_{l\in p}s_l^z(T+\frac{1}{2})\text{ and } d^z(T+\frac{1}{2})=r^z_p(T)r^z_p(T+1)\prod_{l\in p}s_l^x(T+\frac{1}{2}).
\end{equation}
Therefore, to preserve the parity of the detectors, the reparameterization in Eq.~\eqref{eq:reparam} between $t=T$ and $t=T+1$ is modified as
\begin{align}
    &s^{x}_l(T+\frac{1}{2})\to s^{x}_l(T+\frac{1}{2})\sigma^x_l(T)\sigma^x_l(T+1)\sigma^x_{l_1}(T+\frac{1}{2})\sigma^x_{l_2}(T+\frac{1}{2}),\nonumber\\
    &s^{z}_l(T+\frac{1}{2})\to s^{z}_l(T+\frac{1}{2})\sigma^x_{\bar l}(T)
    \sigma^z_l(T)\sigma^z_l(T+1)\sigma^z_{l_1}(T+\frac{1}{2})\sigma^z_{l_2}(T+\frac{1}{2}).
\end{align}
Again, $\bar p$ and $\bar l$ are $p$ and $l$ under mirror-reflection with respect to the diagonal line, respectively. In the stat-mech model in Eq.~\eqref{eq:fullstatmech}, the terms involving $s^{z}_l$ will be modified in the same way as the RHS above.

We now discuss the impact of the t$S$ gate on the error threshold using the stat-mech model. Again, we assume the independent-error model, in which $X$ and $Z$ errors occur independently, so that $J_z=0$ in the Hamiltonian. Therefore, the two flavors of Ising spins couple only through terms involving $s^z_l(T+\frac{1}{2})$. Since the disorder distributions are translation invariant, we can mirror-reflect the spatial location of all the $\sigma^x$ spins, so that the couplings are geometrically local. We recognize that, the stat-mech model in this case is essentially the same as Eq.~\eqref{eq:planedef} for two toric code blocks undergoing a tCNOT gate, where the two spins for the control and target code blocks become $\{\sigma^x\}$ and $\{\sigma^z\}$ spins for $X$ and $Z$ errors, and the bit-flip error rates of the two code blocks become the bit-flip and phase-flip error rate of the single code block. Therefore, for the t$S$ gate, the error threshold of logical $Z$ error will be locally lowered near $t=T$, similar to the target block of the tCNOT gate. In the stat-mech model, this effect can be effectively captured by the increase in local temperature of the $\sigma^z$ spins, as sketched in Fig.~\ref{fig:timeline}.

\section{Stat-mech mapping for generic CSS codes under transversal Clifford gates}
\label{app:general}

In this section, we generalize the stat-mech mapping approach to generic Calderbank-Steane-Shor (CSS) codes that support global transversal Clifford gates.

\subsection{Transversal logical circuit in the binary matrix formalism}

A CSS code over $n$ qubits can be defined by two binary parity check matrices $H_X\in \{0,1\}^{n_x\times n}$ and $H_Z\in \{0,1\}^{n_z\times n}$. Every row in $H_{X/Z}$ represents an $X/Z$ check, and $n_{x/z}$ are numbers of $X/Z$ checks in the code, respectively. Every Pauli error can be parameterized by a pair of binary vectors $e\equiv(\vec{e}_z,\vec{e}_x)^T$, where $\vec{e}_{x,z}\in \{0,1\}^{n}$ indicates the absence/existence of a Pauli $X$ or $Z$ error at each of the $n$ qubits. Assuming perfect syndromes, the Pauli error $e$ triggers a syndrome 
\begin{equation}
    d=He\equiv \left(\begin{array}{cc}
         H_X&0  \\
         0& H_Z
    \end{array}\right)\left(\begin{array}{cc}
        \vec{e}_z\\
         \vec{e}_x 
    \end{array}\right),
\end{equation}
where $d=(d_x,d_z)^T$ and $s_{x/z}\in\{0,1\}^{n_{x/z}}$ are $X$ and $Z$ syndromes, respectively. A $Z$ logical operator is represented by a binary vector $l_z=(\vec{l},0)$ where $\vec{l}\in \ker{H_X}/\text{rs}(H_Z)$, ($\text{rs}(H_Z)$ is the row space generated by the rows of $H_Z$). Similarly, an $X$ logical operator is represented by $l_x=(0,\vec{l}_x)$ where  $\vec{l}_x\in \ker{H_Z}/\text{rs}(H_X)$.

The error-detection mechanism of a CSS code under repeated noisy syndrome extractions can also be understood using an extended version of the binary matrix formalism. Here, Pauli and syndrome errors over spacetime can be parameterized by a single binary vector $\mathfrak{e}$ over time\footnote{In this section, the Fraktur alphabet always denotes the binary vector representation of a spacetime error, while the vector notation is used for binary vectors of length $n$, i.e. the physical errors.}:
\begin{equation}
    \mathfrak{e}\equiv \left(\begin{array}{c >{\color{blue}}c >{\color{blue}}c|cc|>{\color{blue}}c >{\color{blue}}c|ccc}
        \cdots & r_x(t) &r_z(t) &\vec{e}_z(t+\frac{1}{2}) &\vec{e}_x(t+\frac{1}{2}) &r_x(t+1), &r_z(t+1) &\vec{e}_z(t+\frac{3}{2}) &\vec{e}_x(t+\frac{3}{2}) &\cdots
    \end{array}\right)^T.
\end{equation}
Here $r_{x/z}(t)\in\{0,1\}^{n_{x/z}}$ represents the absence or existence of the syndrome extraction error at each $X/Z$ check at time $t$. The vertical lines in $\mathfrak{H}$ separate the columns that correspond to the physical Pauli error and syndrome extraction errors, and the convention will be used in subsequent discussions. Hence, the spacetime detectors formed by repeated syndrome measurements, whose parities are represented by a binary vector $\mathfrak{d}$, can be written in the following matrix form:
\begin{equation}
\begin{aligned}
    \mathfrak{d}&\equiv \left(\begin{array}{cccccc}
         \cdots & d^x(t+\frac{1}{2}) & d^z(t+\frac{1}{2})& d^x(t+\frac{3}{2}) & d^z(t+\frac{3}{2}) & \cdots 
    \end{array}\right)^T\\
    &=\mathfrak{H}\mathfrak{e}=\left(\begin{array}{c >{\color{blue}}c>{\color{blue}}c|cc|>{\color{blue}}c>{\color{blue}}c|cc|>{\color{blue}}c>{\color{blue}}cc}
       \ddots & \vdots &\vdots &\vdots &\vdots &\vdots &\vdots &\vdots &\vdots &\vdots &\vdots &\iddots \\
       \cdots  & \mathds{1}_{n_x} & 0 & H_X & 0 & \mathds{1}_{n_x} &0 & 0 & 0 & 0 & 0 & \cdots \\
       \cdots & 0  & \mathds{1}_{n_z} & 0 & H_Z & 0 & \mathds{1}_{n_z} & 0& 0 & 0 & 0& \cdots  \\
       \cdots  & 0 & 0 & 0 & 0 & \mathds{1}_{n_x} &0 & H_X & 0 &\mathds{1}_{n_x} & 0 &\cdots \\
       \cdots &0  & 0 & 0 & 0 & 0 & \mathds{1}_{n_z} &0 & H_Z & 0 & \mathds{1}_{n_z} & \cdots\\
        \iddots & \vdots &\vdots &\vdots &\vdots &\vdots &\vdots &\vdots &\vdots &\vdots &\vdots &\ddots \\
    \end{array}\right)\left(\begin{array}{c}
        \vdots\\ r_x(t)\\r_z(t)\\\vec{e}_z(t+\frac{1}{2})\\\vec{e}_x(t+\frac{1}{2})\\r_x(t+1)\\r_z(t+1)\\\vec{e}_z(t+\frac{3}{2})\\\vec{e}_x(t+\frac{3}{2})\\r_x(t+2)\\r_z(t+2)\\ \vdots
    \end{array}\right),
\end{aligned}
\end{equation}
where $\mathfrak{d}$ collects the parity of the spacetime detectors, $\mathds{1}_{n_{x/z}}$ is the identity matrix of dimension $n_{x/z}$, and $d^{x/z}(t+\frac{1}{2})\in\{0,1\}^{n_{x/z}}$ represents the parity of the $X/Z$ detectors. We will refer to $\mathfrak{H}$ as the \textit{spacetime parity check matrix}. 

We now consider the effect of transversal gates on the detectors. The physical implementation of a transversal logical Clifford gate $\bar U$ in the CSS code can be represented by a binary symplectic matrix $U\in Sp(2n,\mathbb{F}_2)$ which satisfies
\begin{equation}
    U\Omega U^T=\Omega,\ \Omega\equiv\left(\begin{array}{cc}
        \mathds{1}_{n} & 0 \\
        0 & \mathds{1}_n 
    \end{array}\right).
\end{equation}
Due to the transversality of $U$, the stabilizer group of the CSS code should remain the same before and after the transversal gate. Written in binary matrix language, this condition reads
\begin{equation}\label{eq:tcondition}
    HU=V_UH,
\end{equation}
where $V_U\in SL(n_x+n_z,\mathbb{F}_2)$ represents the change each individual stabilizer under the transversal gate. In the detector error model, $V$ determines the detector geometry across the transversal gate. Consider a transversal gate $U$ applied between $t=T$ and $t=T+1$ rounds of syndrome extraction. Assuming the physical Pauli errors do not persist across the transversal gate $U$, the spacetime parity check matrix can be written in the following block form:
\begin{equation}\label{eq:ut}
    \mathfrak{H}|_{\bar U(T+\frac{1}{2})}=\left(\begin{array}{c>{\color{blue}}c|c|>{\color{blue}}c|c|>{\color{blue}}cc}
       \ddots & \vdots &\vdots  &\vdots &\vdots &\vdots &\iddots \\
       \cdots  & V_U^{-1} & H & \mathds{1}_{n_x+n_z} & 0 & 0& \cdots \\
       \cdots  & 0 & 0 & \mathds{1}_{n_x+n_z} & H& \mathds{1}_{n_x+n_z} & \cdots \\
        \iddots & \vdots &\vdots  &\vdots &\vdots &\vdots &\ddots \\
    \end{array}\right),
\end{equation}
where $V_U^{-1}$ is located at the block column at time $t=T$. Using Eq.~\eqref{eq:tcondition}, the new spacetime parity check matrix can be related to $\mathfrak{H}$ as
\begin{equation}\label{eq:vuh}
    \mathfrak{v}_U\mathfrak{H}|_{\bar U(T+\frac{1}{2})}=\mathfrak{H}\mathfrak{W}|_{\bar U(T+\frac{1}{2})},\ \mathfrak{v}_U= \text{diag}\left(\cdots,\mathds{1}_{n_x+n_z},V_U,V_U,\cdots\right),\ \mathfrak{W}|_{\bar U(T+\frac{1}{2})}=\text{diag}\left(\cdots,\textcolor{blue}{\mathds{1}_{n_x+n_z}},U,\textcolor{blue}{V_U},U,\textcolor{blue}{V_U},\cdots\right).
\end{equation}
Here $\mathfrak{v}_U$ is a block-diagonal matrix with $V_U$ starting from the row block that represents the detectors between $t=T$ and $t=T+1$ (i.e. the row where $V_U^{-1}$ is at in Eq.~\eqref{eq:ut}), and $\mathds{1}_{n_x+n_z}$ before that. $\mathfrak{W}|_{\bar U(T+\frac{1}{2})}$ is a block-diagonal matrix with alternating $U$ and $V_U$ blocks starting from the columns that represent the physical errors at $t=T+\frac{1}{2}$ (i.e. the column on the right of $V_U^{-1}$ in Eq.~\eqref{eq:ut}).

We note that the binary matrix formalism can be straightforwardly generalized to include multiple transversal gates. For example, if another transversal gate $U'$ is applied at a later time $t=T'>T$, the matrix $\mathfrak{W}$ would become
\begin{align}\label{eq:wmatmultiply}
    \mathfrak{W}|_{\bar U(T+\frac{1}{2})\bar U'(T'+\frac{1}{2})}&= \text{diag}\left(\cdots,\textcolor{blue}{\mathds{1}_{n_x+n_z}},U,\textcolor{blue}{V_U},\cdots, UU', \textcolor{blue}{V_UV_{U'}} ,\cdots\right)\nonumber\\
    &=\text{diag}\left(\cdots,\textcolor{blue}{\mathds{1}_{n_x+n_z}},U,\textcolor{blue}{V_U},\cdots\right)\text{diag}\left(\cdots,\textcolor{blue}{\mathds{1}_{n_x+n_z}},U',\textcolor{blue}{V_{U'}},\cdots\right)=\mathfrak{W}|_{\bar U(T+\frac{1}{2})}\mathfrak{W}|_{\bar U'(T'+\frac{1}{2})}.
\end{align}
Therefore, every time a new transversal gate is applied, $\mathfrak{W}$ is modified by right-multiplication of the corresponding block-diagonal matrix of the transversal gate.

\subsection{Stat-mech mapping via the binary matrix formalism}
Using the binary matrix formalism, we can now discuss the stat-mech mapping with a transversal Clifford gate in the most general setting. Starting from the case of quantum memory. An error model in this case can be specified via the probability distribution
\begin{equation}
    \prob:\mathfrak{e}\mapsto\prob(\mathfrak{e})\in \mathbb{R}_{\geq0}.
\end{equation}
Two spacetime errors $\mathfrak{e}$ and $\mathfrak{e}'=\mathfrak{e}+\delta\mathfrak{e}$ trigger the same set of detectors if they differ by a cycle $\delta\mathfrak{e}\in \ker(\mathfrak{H})$. The trivial cycles, which do not lead to logical errors, are generated by the following four sets of bitstrings
\begin{align}\label{eq:elementarytrivial}
    &\mathfrak{s}^x_i(t+\frac{1}{2})=\left(\begin{array}{c>{\color{blue}}c|cc|>{\color{blue}}cc}
        \cdots & 0 & (H_Z)_i & 0 & 0 & \cdots
    \end{array}\right)^T, \ i\in\{1,2,\dots,n_z\},\nonumber\\
    &\mathfrak{s}^z_i(t+\frac{1}{2})=\left(\begin{array}{c>{\color{blue}}c|cc|>{\color{blue}}cc}
        \cdots & 0 &0 & (H_X)_i & 0 & \cdots
    \end{array}\right)^T, \ i\in\{1,2,\dots,n_x\},\nonumber\\
    &\mathfrak{r}^x_i(t)=\left(\begin{array}{c>{\color{blue}}c|cc|>{\color{blue}}c>{\color{blue}}c|cc|>{\color{blue}}cc}
        \cdots & 0 & \vec{e}_i & 0 & H_X\vec{e}_i & 0 & \vec{e}_i &0 &0 & \cdots
    \end{array}\right)^T, \ i\in\{1,2,\dots,n\},\nonumber\\
    &\mathfrak{r}^z_i(t)=\left(\begin{array}{c>{\color{blue}}c|cc|>{\color{blue}}c>{\color{blue}}c|cc|>{\color{blue}}cc}
        \cdots & 0 & 0 & \vec{e}_i & 0 & H_Z\vec{e}_i & 0 & \vec{e}_i &0 & \cdots
    \end{array}\right)^T, \ i\in\{1,2,\dots,n\}.
\end{align}
Here, the first two sets of cycles, $\mathfrak{s}^{x,z}_i$, are trivial cycles of physical Pauli errors which form a stabilizer themselves. The latter two sets of cycles $\mathfrak{r}^{x,z}_i$, represent physical Pauli errors that are undetected due to syndrome errors, but subsequently occur again and cancel themselves. $(\vec{e}_i)_j=\delta_{i,j}$ is a basis vector that represents a single physical Pauli error. 

Meanwhile, an undetected logical $Z$ or $X$ error can always be represented as
\begin{equation}
    \mathfrak{l}_z=\left(\begin{array}{c>{\color{blue}}c|c|>{\color{blue}}cc}
        \cdots & 0 & l_z  &0 & \cdots
    \end{array}\right)^T, \     \mathfrak{l}_x=\left(\begin{array}{c>{\color{blue}}c|c|>{\color{blue}}cc}
        \cdots & 0 & l_x  &0 & \cdots
    \end{array}\right)^T,
\end{equation}
where $l_{z,x}$ are binary vectors that represent logical operators of the original CSS code\footnote{It does not matter which time the logical error $l_{z,x}$ is located in $\mathfrak{l}_{x,z}$. In fact, using linear combinations of $\mathfrak{r}^{x,z}$, we can always shift it in time.}. In this way, every cycle $\delta \mathfrak{e}$ can be parameterized in the following way:
\begin{equation}
    \delta\mathfrak{e}=\mathfrak{c}(\{\sigma\})+\mathfrak{l},
\end{equation}
where $\mathfrak{l}$ is a linear combination of the logical errors $\mathfrak{l}_{x,z}$, and 
\begin{equation}
    \mathfrak{c}(\{\sigma\})=\sum_t\left\{\sum_{i=1}^n\left[\sigma^x_i(t)\mathfrak{r}^x_i(t)+\sigma^z_i(t)\mathfrak{r}^z_i(t)\right]+\sum_{i=1}^{n_z}\sigma^x_i(t+\frac{1}{2})\mathfrak{s}^x_i(t+\frac{1}{2})+\sum_{i=1}^{n_x}\sigma^z_i(t+\frac{1}{2})\mathfrak{s}^z_i(t+\frac{1}{2})\right\}
\end{equation}
represents a trivial cycle formed by a linear combination of the four sets of elementary trivial cycles in Eq.~\eqref{eq:elementarytrivial}. $\{\sigma^{x,z}_i(t)=0,1\}$ and $\{\sigma^{x,z}_i(t+\frac{1}{2})=0, 1\}$ are \textit{binary} spins. In the context of toric code under noisy syndromes, they correspond to the value $\frac{1-\sigma^{x,z}}{2}$ of the $\sigma^{x,z}$ spins at the respective times in Fig.~\ref{fig:3dsm}.

Putting everything together and using the definition in Eq.~\eqref{eq:logicalprob}, the logical error rate of the decoder is
\begin{equation}
    P_\text{logical}=\sum_{\mathfrak{e}}\prob(\mathfrak{e})\frac{\sum_{\mathfrak{l}\neq0,\{\sigma\}}\prob(\mathfrak{e+\mathfrak{l}+\mathfrak{c}(\{\sigma\})})}{\sum_{\mathfrak{l},\{\sigma\}}\prob(\mathfrak{e+\mathfrak{l}+\mathfrak{c}(\{\sigma\})})}.
\end{equation}
Therefore, the partition function of the stat-mech model reads
\begin{equation}
    Z(\mathfrak{e})\propto\sum_{\{\sigma\}}\prob(\mathfrak{e}+\mathfrak{c}(\{\sigma\})).
\end{equation}
The quenched disorder, $\mathfrak{e}$, follows the distribution $\prob(\mathfrak{e})$. For structured CSS codes (such as topological codes) and a distribution of errors $\prob(\mathfrak{e})$, the partition function can be brought into the Boltzmann form of a known Hamiltonian.

Now consider the case with a transversal logical gate. From Eq.~\eqref{eq:vuh}, we see that every trivial cycle $\mathfrak{c}'$ of the new spacetime parity check matrix $\mathfrak{H}'$ with transversal gates is related to the kernel of $\mathfrak{H}$ via $\mathfrak{c}'=\mathfrak{W}^{-1}\mathfrak{c}$, $\mathfrak{c}\in \ker(\mathfrak{H})$, where $\mathfrak{W}$ can be computed using Eq.~\eqref{eq:wmatmultiply}. Therefore, the new partition function reads
\begin{equation}
    Z'(\mathfrak{e})\propto\sum_{\{\sigma\}}\prob(\mathfrak{e}+\mathfrak{W}^{-1}\mathfrak{c}(\{\sigma\})).
\end{equation}
Since the $\sigma$ spins are summed over, one may wonder if the new partition function can be brought to $Z(\mathfrak{e})$ via redefinition of the spins, which would undo the unitary $\mathfrak{W}^{-1}$ in $Z'(\mathfrak{e})$. This turns out to be \textit{impossible}. To see this, we show that 
\begin{equation}\label{eq:kerneq}
    \ker(\mathfrak{H})\neq \ker(\mathfrak{H}')
\end{equation}
for a single transversal logical gate $\bar U$, in which case $\mathfrak{H}'=\mathfrak{H}_{\bar U(T+\frac{1}{2})}$. The two kernels are spanned by $\{\mathfrak{s}^{x,z}_i(t+\frac{1}{2}),\mathfrak{r}^{x,z}_i(t)\}$ and $\{\mathfrak{W}^{-1}\mathfrak{s}^{x,z}_i(t+\frac{1}{2}),\mathfrak{W}^{-1}\mathfrak{r}^{x,z}_i(t)\}$, respectively. We observe that:
\begin{enumerate}
    \item The cycles $\mathfrak{r}^{x,z}_i(t)$ are linearly independent for every $i$ and $t$;
    \item $\mathfrak{W}^{-1}\mathfrak{r}^{x}_i(T)=\left(\begin{array}{c>{\color{blue}}c|cc|>{\color{blue}}c>{\color{blue}}c|c|>{\color{blue}}cc}
        \cdots & 0 & \vec{e}_i & 0 & H_X\vec{e}_i & 0 & U^{-1}(\vec{e}_i, 0) &0 & \cdots
    \end{array}\right)^T$, and similarly for $\mathfrak{W}^{-1}\mathfrak{r}^{z}_i(T)$. They differ from $\mathfrak{r}^{x,z}_i(T)$ by the entries to the physical errors at $T+\frac{1}{2}$.
\end{enumerate}
Therefore, the new cycles $\mathfrak{W}^{-1}\mathfrak{r}^{x,z}_i(T)$ cannot be obtained from linear combinations of the original cycles $\mathfrak{r}^{x,z}_i(T)$. This proves Eq.~\eqref{eq:kerneq}.

Nevertheless, the two partition functions differ only locally near $t=T$. In fact, away from the transversal gate (i.e. when $t\geq T+1$), it is easy to verify that the new cycles $\mathfrak{W}^{-1}\mathfrak{r}^{x,z}_i(t\geq T
)$ are linear combinations of the old ones:
\begin{equation}
    \left(\begin{array}{ccc|ccc}
        \mathfrak{W}^{-1}\mathfrak{r}^x_1(t) & \cdots & \mathfrak{W}^{-1}\mathfrak{r}^x_n(t) & \mathfrak{W}^{-1}\mathfrak{r}^z_1(t) &\cdots & \mathfrak{W}^{-1} \mathfrak{r}^z_n(t)
    \end{array}\right) =U^{-1} \left(\begin{array}{ccc|ccc}
        \mathfrak{r}^x_1(t) & \cdots & \mathfrak{r}^x_n(t) & \mathfrak{r}^z_1(t) &\cdots &  \mathfrak{r}^z_n(t)
    \end{array}\right) .
\end{equation}
Hence, at $t\geq T$, $Z(\mathfrak{e}')$ can be brought to $Z(\mathfrak{e})$ via redefining the $\{\sigma\}$ spins. In other words, the impact of every transversal gate on the stat-mech model is always local at the time it is implemented.

\end{document}